\documentclass[aps,pre,twocolumn,superscriptaddress,secnumarabic,nofootinbib]{revtex4-2}

\usepackage[total={7.2in,9in}]{geometry}
\usepackage{amsmath, amssymb, amsfonts}
\usepackage{graphicx}
\usepackage[colorlinks, linkcolor=blue, citecolor=blue, urlcolor=blue]{hyperref}
\usepackage{bm}
\usepackage{physics}
\usepackage[version=4]{mhchem}
\usepackage[utf8]{inputenc}
\usepackage{geometry}
\usepackage{amsmath}
\usepackage{fancyhdr}
\usepackage{graphicx}
\usepackage{framed}
\usepackage{amsthm, amssymb, appendix, bm, graphicx, mathrsfs}
\usepackage{orcidlink}
\usepackage{subcaption}

\begin{document}

\title{Bell–Plesset Effects on Rayleigh Taylor Instability of Three Dimensional Spherical Geometry}

\author{Xilai Li\orcidlink{0009-0004-4125-1057}}
\email{martianian@sjtu.edu.cn}
\thanks{Equal contribution.}
\affiliation{Zhiyuan College, Shanghai Jiao Tong University, Shanghai, 201210, China}

\author{Yilin Wu\orcidlink{0009-0004-6551-1853}}
\email{YilinWu2026@gmail.com}
\thanks{Equal contribution.}
\affiliation{Zhiyuan College, Shanghai Jiao Tong University, Shanghai, 201210, China}

\author{Zhengnuo Chen\orcidlink{0009-0007-6376-8606}}
\email{chenzhengnuo@sjtu.edu.cn}
\thanks{Equal contribution.}
\affiliation{Zhiyuan College, Shanghai Jiao Tong University, Shanghai, 201210, China}

\author{Mengqi Yang\orcidlink{0009-0007-2022-721X}}
\affiliation{Tsung-Dao Lee Institute, Shanghai Jiao Tong University, Shanghai, 201210, China}
\affiliation{Key Laboratory for Laser Plasmas (MoE) and School of Physics and Astronomy, Shanghai Jiao Tong University, Shanghai 200240, China}
\affiliation{Collaborative Innovation Center of IFSA, Shanghai Jiao Tong University, Shanghai 200240, China}

\author{Jie Zhang\orcidlink{0000-0001-7821-4808}}
\affiliation{Tsung-Dao Lee Institute, Shanghai Jiao Tong University, Shanghai, 201210, China}
\affiliation{Key Laboratory for Laser Plasmas (MoE) and School of Physics and Astronomy, Shanghai Jiao Tong University, Shanghai 200240, China}
\affiliation{Collaborative Innovation Center of IFSA, Shanghai Jiao Tong University, Shanghai 200240, China}

\date{\today}

\begin{abstract}
We develop a weakly nonlinear, multi-mode theory for the Rayleigh–Taylor instability (RTI) on a time-varying spherical interface, fully incorporating mode couplings and the Bell–Plesset (BP) effects arising from interface convergence. Our model extends prior analyses, which have been largely restricted to static backgrounds, 2D cylindrical geometries, or single-mode initial condition. We present a framework capable of evolving arbitrary, fully three-dimensional initial perturbations on a dynamic background. At the first order, mode amplitudes respond to the time-varying interface acceleration with an exponential-like growth, in qualitative agreement with classic static results. At second order, nonlinear mode coupling reveals a powerful selection rule: energy is preferentially channeled into axisymmetric ($m=0$) modes. We find that the BP effects dramatically amplify the instability growth by a few orders of magnitude, with this amplification being even more significant for second order couplings. Despite this strong channeling, the second order amplitudes remain small relative to the first order, validating the perturbative approach. These findings offer new physical insights into time-dependent interface instabilities relevant to applications such as astrophysical shell collapse and inertial confinement fusion, highlighting the uniquely dominant role of axisymmetric modes in BP-driven convergent flows.

\end{abstract}

\maketitle

\section{Introduction}
The Rayleigh Taylor instability (RTI)~\cite{RTI1,RTI2} is a fundamental fluid instability that arises at the interface between two fluids of different densities when the lighter fluid supports the heavier one. RTI is a critical mechanism in a wide range of physical systems, from astrophysical phenomena such as supernova explosions and nebula formation~\cite{SNe1,SNe2,Nebula1,Nebula2} to technological applications like inertial confinement fusion (ICF)~\cite{Hurricane1,Hurricane2,Hurricane3}, and meteorological flows~\cite{meteorology1,meteorology2,meteorology3}. While the linear stage of RTI is well understood, its dynamics become substantially more complex in the nonlinear regime. Weakly nonlinear (WN) analyses and specialized experiments under idealized conditions have been instrumental in revealing the rich evolutionary behaviors that characterize this stage~\cite{RTI_WN1,RTI_WN2}, including characteristic structures of penetrating bubbles and spikes~\cite{bubble_spike1, bubble_spike2, bubble_spike3}.

Analytic investigations of RTI typically employ perturbative expansions to describe the evolution of the fluid interface~\cite{RTI_2d_MultiMode, perturbation1, perturbation2, perturbation3, RTI_3d_spherical2}. In a static spherical geometry, the interface radius and the velocity potentials of the two fluids are expanded in spherical harmonics \(Y_{l,m}\). The first-order solution shows that each mode grows exponentially and independently, with a characteristic growth rate
\begin{equation}
\gamma_l = \sqrt{\frac{l(l+1)(\rho_{\rm ex} - \rho_{\rm in})\, g}{\left[(l+1)\rho_{\rm in} + l\rho_{\rm ex}\right]R}}.
\label{eq:reduced_growth_factor}
\end{equation}
At higher orders, nonlinear mode coupling emerges, and the strongest growth typically occurs for azimuthal indices near \(m \simeq (l+1)/2\)~\cite{RTI_3d_spherical1, RTI_3d_spherical2}. 
However, previous studies compute the higher order evolution only for single-mode initial perturbations and therefore cannot describe the RTI dynamics for an arbitrary interface shape. Moreover, these analyses are fundamentally altered in realistic settings where RTI develops on a dynamic background and Bell-Plesset (BP) effects become significant.

The complexity of RTI is amplified in converging geometries, such as those found in imploding ICF capsules or the collapsing cores of massive stars. In these systems, the time-dependent curvature and background compression introduce the BP effects, which profoundly alters the instability's growth rates and nonlinear behaviors~\cite{BPe1,BPe2,BPe3}. Previous studies of cylindrical implosions (equivalently 2D spherical geometry) have shown that under such conditions, mode coupling emerges beyond the first order approximation, fundamentally altering the RTI dynamics~\cite{Zhang2024}. Moreover, BP effects introduce considerable complexities into the evolution equations, precluding closed-form analytic solutions. Although extensive research has explored BP-modified RTI in cylindrical geometries~\cite{BPe4,BPe5,BPe6,BPe7}, a comprehensive understanding of these effects on a fully three-dimensional spherical interface remains a formidable challenge, primarily due to the high dimensionality of the problem.

Motivated by these considerations, this paper presents a theoretical investigation of RTI development in a fully three-dimensional spherical convergent flow. We carry the perturbation expansion to second order to systematically examine the interplay between Bell--Plesset effects and multi-mode interactions. We organize this paper as follows. Section~\ref{sec:equations} introduces our theoretical framework and fundamental equations. Section~\ref{sec:numerical} describes the setup and numerical methods to solve the equations. Section~\ref{sec:results_discussions} presents the evolution of first-order (\ref{subsec:first_order_results}) and second-order interface perturbations for single-mode and uniformly distributed multimode initial perturbations (\ref{subsec:second_order_results}), “bubble”-like initial conditions (\ref{subsec:bubble_results}), and compares results with and without BP effects~\ref{subsec:bp_results}. Finally, Section~\ref{sec:conclusions} summarizes our conclusions and discusses the broader implications.

\section{Theoretical framework}
\label{sec:equations}

We investigate the spherical Rayleigh--Taylor instability in the presence of geometric compression. Owing to the intrinsic complexity of boundary-layer dynamics, particular care is required in distinguishing compressible and incompressible effects and in identifying the spatial and temporal scales on which each description applies. Within this controlled framework, the model can yield physically instructive results under a clearly defined set of assumptions.

The analysis is formulated in a spherical coordinate system, where \( r \), \( \theta \), and \( \varphi \) denote the radial, poloidal, and azimuthal coordinates, respectively. The interface between the two fluids is described by a perturbed spherical surface,
\begin{equation}
    r_{\mathrm{sf}}(\theta, \varphi, t) = R(t) + \eta(\theta, \varphi, t),
\end{equation}

where \( R(t) \) is the unperturbed radius and \( \eta(\theta, \varphi, t) \) denotes the interfacial perturbation. We restrict attention to perturbations confined to a thin spherical shell,
\begin{equation}
    |\eta(\theta, \varphi, t)| \ll R(t),
\end{equation}
which defines the regime of validity of the present theory.

The fluid is assumed to be irrotational, allowing the velocity field to be expressed in terms of a scalar potential,
\begin{equation}
    \Phi^i(r,\theta,\varphi) = \psi(r) + \phi^i(r,\theta,\varphi),
\end{equation}
where the superscript \( i \in \{\mathrm{in},\mathrm{ex}\} \) labels the interior and exterior regions. The background potential \( \psi \) accounts for geometric effects associated with the compression of the sphere and depends only on the radial coordinate, reflecting compressibility at the global scale. The perturbative potentials \( \phi^i \) are associated with the perturbation velocity, which is assumed to induce negligible density variations, i.e., $|\nabla^2\phi^i|\ll|\nabla^2\psi|$. Without loss of generality, we further assume the perturbation velocity to be strictly incompressible, such that $\nabla^2\phi^i=0$.

We focus on a thin spherical shell at radius $R$ that contains the interface of interest. Within this thin layer, the density on each side of the interface is assumed to be homogeneous. We further introduce a local density compression rate for the thin layer,
\begin{equation}
    \gamma_\rho \equiv \frac{\dot{\rho}}{\rho}\bigg|_{\text{thin layer}}
    \equiv -\frac{\dot{V}}{V}\bigg|_{\text{thin layer}}
    = -\frac{3}{r^3 - R^3}\left(r^2 \dot{r} - R^2 \dot{R}\right),
    \label{eq:gamma_rho_def}
\end{equation}
where \( \rho \) is the density of this layer, and \( r \) denotes an arbitrary radius near \( R \). Note that $\gamma_\rho$ is independent of the choice of $r$, provided it remains a small perturbation from $R$, due to the homogeneity assumption.

As the density compression effect is caused by the background convergent flow \( \psi \), thus,
\begin{equation}
    \frac{\partial \psi}{\partial r}
    \equiv \dot{r}\big|_{\text{BG}}
    = \frac{R^2 \dot{R}}{r^2}
      + \frac{\gamma_\rho}{3}\left(\frac{R^3}{r^2} - r \right),
    \label{eq:psi_by_gamma_rho}
\end{equation}
with $\dot{r}\big|_{\text{BG}}$ being the velocity of background (BG) flow.

The governing equations follow from the incompressibility of the perturbative flow, together with continuity of velocity and pressure across the interface. Spatial derivatives of the density are neglected due to the thin-layer approximation, while temporal derivatives are omitted because density evolution is governed by the background convergent flow. The resulting equations are
\begin{equation}
    \nabla^2 \phi^i = 0,
    \label{eq:laplace}
\end{equation}
\begin{equation}
\begin{split}
    \frac{\partial \eta}{\partial t}
    &+ \frac{1}{r^2\sin^2\theta}
      \frac{\partial\eta}{\partial \varphi}
      \frac{\partial\phi^i}{\partial \varphi}
    + \frac{1}{r^2}
      \frac{\partial\eta}{\partial \theta}
      \frac{\partial\phi^i}{\partial \theta} \\
    &- \frac{\partial\phi^i}{\partial r}
    + \left(\dot{R} - \frac{\partial \psi}{\partial r} \right)
    = 0 \quad \text{at } r = R + \eta,
    \label{eq:velocity_continuity}
\end{split}
\end{equation}

\begin{equation}
\begin{split}
    \rho^{i}\biggl\{
    &\frac{1}{2}\bigg[
    \frac{1}{r^2\sin^2\theta}
    \left(\frac{\partial \Phi^i}{\partial \varphi}\right)^2
    + \frac{1}{r^2}
    \left(\frac{\partial \Phi^i}{\partial \theta}\right)^2
    + \left(\frac{\partial \Phi^i}{\partial r}\right)^2
    \bigg] \\
    &+ \frac{\partial \Phi^i}{\partial t}
    + g r
    \biggr\}
    \Bigg|^{i=\mathrm{ex}}_{i=\mathrm{in}}
    = f(t)
    \quad \text{at } r = R + \eta,
    \label{eq:pressure_continuity}
\end{split}
\end{equation}
where \( f(t) \) is an arbitrary function of time, which is hereafter set to zero, and \( g \) denotes the external acceleration. Despite this numerical equivalence, $g$ is conceptually distinct from $\ddot{R}$: it represents the external force driving the instability rather than a geometrical effect. In the classical literature, $g$ typically appears as gravitational acceleration; in our setting, however, external driving forces dominate and gravity can be neglected. We retain it here to facilitate a visual comparison with the case without a convergent background flow (the BP effects, see Section~\ref{subsec:bp_results}).

The perturbations are expanded in spherical harmonics,
\begin{equation}
    \eta(\theta, \varphi, t)
    = \sum_{l,m} a_{l,m}(t) Y_l^m(\theta, \varphi),
\end{equation}
\begin{equation}
    \phi^{\mathrm{ex}}(r,\theta,\varphi,t)
    = \sum_{l,m} b^{\mathrm{ex}}_{l,m}(t)
    \left(\frac{r}{R}\right)^{-(l+1)} Y_l^m(\theta,\varphi),
\end{equation}
\begin{equation}
    \phi^{\mathrm{in}}(r,\theta,\varphi,t)
    = \sum_{l,m} b^{\mathrm{in}}_{l,m}(t)
    \left(\frac{r}{R}\right)^l Y_l^m(\theta,\varphi),
\end{equation}
which automatically satisfy Eq.~\eqref{eq:laplace} and the boundary conditions at \( r=0 \) and \( r\to\infty \).

Considering there exist many kinds of definitions of $Y_l^m$,our definition is
\begin{equation}
    Y_l^m(\theta, \varphi) = \sqrt{\frac{2l+1}{4\pi}\frac{(l-m)!}{(l+m)!}} P_l^m(\cos\theta) e^{im\varphi},
\end{equation}
with $P_l^m$ being the associated Legendre polynomials.

The amplitudes are further expanded in a formal small parameter \( \epsilon \),
\begin{equation}
    a_{l,m} = \epsilon a^{(1)}_{l,m}+\epsilon^2a^{(2)}_{l,m} +\cdots,
\end{equation}
\begin{equation}
    b^{\rm{in}}_{l,m} = \epsilon b^{\text{in}(1)}_{l,m}+\epsilon^2b^{\text{in}(2)}_{l,m} +\cdots,
\end{equation}
\begin{equation}
    b^{\rm{ex}}_{l,m} = \epsilon b^{\text{ex}(1)}_{l,m}+\epsilon^2b^{\text{ex}(2)}_{l,m} +\cdots.
\end{equation}
The separation by powers of \( \epsilon \) serves as a bookkeeping device to resolve the governing equations across multiple scales, with finer structure captured by the higher-order terms.

Substituting these expansions into Eqs.~\eqref{eq:velocity_continuity} and \eqref{eq:pressure_continuity} yields a hierarchy of coupled equations, whose explicit forms are given in Appendix~\ref{app:equation_expansion}. In practice, we set \( \epsilon = 1 \) to facilitate comparison between different orders. The resulting equations are solved order by order to obtain the mode amplitudes. The properties of the first- and second-order solutions are discussed below.

\subsection{First Order Equation}

At first order, $b^{\text{in}(1)}_{l,m}$ and $b^{\text{ex}(1)}_{l,m}$ are related to $a^{(1)}_{l,m}$ as
\begin{equation}
    b^{\text{in}(1)}_{l,m} = \frac{R\dot{a}^{(1)}_{l,m} + (2\dot{R} + R\gamma_\rho)a^{(1)}_{l,m}}{l},
    \label{eq:bin1_by_a1}
\end{equation}
\begin{equation}
    b^{\text{ex}(1)}_{l,m} = -\frac{R\dot{a}^{(1)}_{l,m} + (2\dot{R} + R\gamma_\rho)a^{(1)}_{l,m}}{l+1},
    \label{eq:bex1_by_a1}
\end{equation}

Substituting into the momentum conservation equation yields
\begin{equation}
\begin{split}
    \frac{l(l+1)g\left(\rho_\text{in} - \rho_\text{ex} \right) + \left[(l+1)(l+2)\rho_\text{in} - l(l-1)\rho_\text{ex} \right]\ddot{R}}{\left[(l+1)\rho_\text{in} + l\rho_\text{ex} \right]R}a&\\
    + \frac{\left(\gamma_\rho\dot{R} + \dot{\gamma}_\rho R \right)}{R}a+ \left(\gamma_\rho + \frac{3\dot{R}}{R}\right)\dot{a}+\ddot{a} = 0. \quad \quad \quad \quad&
    \label{eq:first_order_eq}
\end{split}
\end{equation}
For notational convenience, we have omitted the mode indices \( l \) and \( m \) on the first-order amplitude \( a \) in this expression. 
The first-order evolution is independent of azimuthal number $m$. In the static limit where there's no BP effects ($R=\text{const}$, $\gamma_\rho=0$), the reduced growth factor is same with Eq.\eqref{eq:reduced_growth_factor}, thus validating our derivations in none-BP limit.

To better understand this result, we simplify Eq.~\eqref{eq:first_order_eq} by defining
\begin{equation}
    \gamma_R=\frac{\dot{R}}{R},\quad u= \frac{2l(l+1)\alpha}{1+2l+\alpha},
    \label{eq:first_order_simplification_parameters}
\end{equation}
where $\alpha=\frac{\rho^\mathrm{in}-\rho^\mathrm{ex}}{\rho^\mathrm{in}+\rho^\mathrm{ex}}$ is the Atwood number \cite{perturbation2,Atwood2}. The evolution Eq.\eqref{eq:first_order_eq} then becomes
\begin{equation}
    \ddot{a}+(\gamma_\rho+3\gamma_R)\dot{a}+\left[\gamma_\rho\gamma_R +\dot{\gamma}_\rho +\frac{ug}{R} +(u+2)(\dot{\gamma}_R+\gamma_R^2) \right]a=0.
    \label{eq:first_order_simplified}
\end{equation}
Let $x(t)$ and $y(t)$ denote the coefficients of $\dot{a}$ and $a$
\begin{equation}
    x(t)=\gamma_\rho+3\gamma_R,
    \label{eq:xt_Def}
\end{equation}
\begin{equation}
    y(t)=\gamma_\rho\gamma_R +\dot{\gamma}_\rho +\frac{ug}{R} +(u+2)(\dot{\gamma}_R+\gamma_R^2),
    \label{eq:yt_def}
\end{equation}
and define
\begin{equation}
    a(t) = p(t)\exp\left[-\frac{1}{2}\int x(t)\mathrm{d}t\right]=\frac{C_0}{\sqrt{\rho R^3}}p(t),
    \label{eq:bt_def}
\end{equation}
where $C_0$ is the integral constant. The equation then simplifies to
\begin{equation}
    \ddot{p} + z(t)p = 0,
\label{eq:first_order_oversimplified}
\end{equation}
where $z(t)=y(t) - \frac{1}{4}\dot{x}(t) - \frac{1}{2}x^2(t)$.

Generally, this equation lacks an analytic solution. The sign of $z(t)$ determines instability: $z<0$ produces exponential-like amplification, while $z>0$ results in sinusoidal fluctuations and thus suppresses the instability. Since it exhibits scale invariance with respect to the initial amplitude and independent from the azimuthal index \(m\), we solve for a normalized growth factor $g_l(t)$ for each mode number \(l\), defined as
\begin{equation}
    g_l(t) = \frac{p_l(t)}{p_l(t_0)} =\frac{a^{(1)}_{l,m}(t)}{a^{(1)}_{l,m}(t_0)},
    \label{eq:first_order_growth_factor_def}
\end{equation}
which is identical for all \(m\) at a given \(l\). This growth factor obeys the same differential form as Eq.~\eqref{eq:first_order_oversimplified}, which is subjected to the initial conditions of a quiescent interface
\begin{equation}
    g_l(t_0) = 1,\quad \dot{g}_l(t_0)=0.
    \label{eq:first_order_growth_factor_initial_condition}
\end{equation}

\subsection{Second Order Equations}

The second-order equations are considerably complex and their full derivations are presented in Appendix~\ref{app:second_order_equation}.

The mass conservation equations for the internal component is given by

\begin{widetext}
\begin{equation}
    \begin{split}
        &\frac{2}{3}\left[\dot{a}^{(2)}_{l,m} + \left(2\gamma_R + \gamma_\rho\right)a^{(2)}_{l,m} -b^{\text{in}(2)}_{l,m}\frac{l}{R} \right]
        + \sum_{l_1,m_1}\bigg[\dot{a}^{(2)}_{l_1,m_1}+ \left(2\gamma_R + \gamma_\rho\right)a^{(2)}_{l_1,m_1} -b^{\text{in}(2)}_{l_1,m_1}\frac{l_1}{R}\bigg] A_{l,m,l_1,m_1} 
        +\frac{1}{R^2}\sum_{l_1,m_1,l_2,m_2} \bigg\{a^{(1)}_{l_1,m_1} b^{\text{in}(1)}_{l_2,m_2} \\ &\quad \times \Theta_{l,m,l_1,m_1,l_2,m_2} - m_1 m_2
        a^{(1)}_{l_1,m_1} b^{\text{in}(1)}_{l_2,m_2}B_{l,m,l_1,m_1,l_2,m_2}
       -\big[3\dot{R} + R\gamma_\rho + l_2(l_2-1)a^{(1)}_{l_1,m_1} b^{\text{in}(1)}_{l_2,m_2}\big] H_{l,m,l_1,m_1,l_2,m_2}  \bigg\} = 0,
    \end{split}
    \label{eq:second_order_eq1_in}
\end{equation}
\end{widetext}

and for the external component we have

\begin{widetext}
\begin{equation}
    \begin{split}
        \frac{2}{3}&\left[\dot{a}^{(2)}_{l,m} + \left(2\gamma_R + \gamma_\rho\right)a^{(2)}_{l,m} +b^{\text{ex}(2)}_{l,m}\frac{l+1}{R} \right]
        + \sum_{l_1,m_1}\left[\dot{a}^{(2)}_{l_1,m_1} + \left(2\gamma_R + \gamma_\rho\right)a^{(2)}_{l_1,m_1} + b^{\text{ex}(2)}_{l_1,m_1}\frac{l_1+1}{R}\right] A_{l,m,l_1,m_1}
        \\ + & \frac{1}{R^2}\sum_{l_1,m_1,l_2,m_2}\bigg\{ a^{(1)}_{l_1,m_1} b^{\text{ex}(1)}_{l_2,m_2} \Theta_{l,m,l_1,m_1,l_2,m_2} - m_1 m_2 a^{(1)}_{l_1,m_1} b^{\text{ex}(1)}_{l_2,m_2}B_{l,m,l_1,m_1,l_2,m_2}  -\\& \big[3\dot{R} + R\gamma_\rho + (l_2+1)(l_2+2)a^{(1)}_{l_1,m_1} b^{\text{ex}(1)}_{l_2,m_2}\big] H_{l,m,l_1,m_1,l_2,m_2} \bigg\} = 0,
    \end{split}
    \label{eq:second_order_eq1_ex}
\end{equation}
\end{widetext}
where $A_{l,m,l_1,m_1}$, $B_{l,m,l_1,m_1,l_2,m_2}$, $H_{l,m,l_1,m_1,l_2,m_2}$, and $\Theta_{l,m,l_1,m_1,l_2,m_2}$ are quantities defined in Appendix~\ref{app:second_order_equation}.

The momentum conservation equation is expressed as

\begin{widetext}
\begin{equation}
    \begin{split}
        \rho^\mathrm{in}& \Bigg\{\frac{2}{3}\left[(\ddot{R}+g){a}^{(2)}_{l,m} + \dot{b}^{\text{in}(2)}_{l,m} \right] + \sum_{l_1,m_1} \bigg[(\ddot{R}+g){a}^{(2)}_{l_1,m_1}+ \dot b^{\text{in}(2)}_{l_1,m_1}\bigg] A_{l,m,l_1,m_1} + \sum_{l_1,m_1,l_2,m_2} \Bigg[\frac{l_2}{R}a^{(1)}_{l_1,m_1} \dot{b}^{\text{in}(1)}_{l_2,m_2} - l_2^2 \frac{\gamma_R}{R} a^{(1)}_{l_1,m_1} b^{\text{in}(1)}_{l_2,m_2} \\
        & + \frac{1}{R^2}\bigg( \dot{R}^2 - R\ddot{R} - \frac{\dot{\gamma}_\rho R^2}{2}\bigg) a^{(1)}_{l_1,m_1} a^{(1)}_{l_2,m_2} + \frac{1}{R}\Big( l_2(l_2 -1)\gamma_R - l_2(2\gamma_R + \gamma_\rho)\Big) a^{(1)}_{l_1,m_1} b^{\text{in}(1)}_{l_2,m_2} + \frac{(2\gamma_R + \gamma_\rho)^2}{2} a^{(1)}_{l_1,m_1} a^{(1)}_{l_2,m_2} \\
        & + \frac{l_1 l_2}{2R^2} b^{\text{in}(1)}_{l_1,m_1} b^{\text{in}(1)}_{l_2,m_2} \Bigg]H_{l,m,l_1,m_1,l_2,m_2} + \frac{1}{2R^2}\sum_{l_1,m_1,l_2,m_2} \bigg( b^{\text{in}(1)}_{l_1,m_1} b^{\text{in}(1)}_{l_2,m_2} \Theta_{l,m,l_1,m_1,l_2,m_2} -m_1m_2b^{\text{in}(1)}_{l_1,m_1}  b^{\text{in}(1)}_{l_2,m_2} B_{l,m,l_1,m_1,l_2,m_2} \bigg) \Bigg\}\\
        &\quad \quad = \\
        \rho^\mathrm{ex}& \Bigg\{\frac{2}{3}\left[(\ddot{R}+g){a}^{(2)}_{l,m} + \dot{b}^{\text{ex}(2)}_{l,m} \right] + \sum_{l_1,m_1} \bigg[(\ddot{R}+g){a}^{(2)}_{l_1,m_1}+ \dot b^{\text{ex}(2)}_{l_1,m_1}\bigg] A_{l,m,l_1,m_1} + \sum_{l_1,m_1,l_2,m_2} \Bigg[-(l_2+1)^2 \frac{\gamma_R}{R}  a^{(1)}_{l_1,m_1} b^{\text{ex}(1)}_{l_2,m_2} \\
        & - \frac{l_2+1}{R}a^{(1)}_{l_1,m_1} \dot{b}^{\text{ex}(1)}_{l_2,m_2} + \frac{1}{R^2}\bigg( \dot{R}^2 - R\ddot{R} - \frac{\dot{\gamma}_\rho R^2}{2}\bigg) a^{(1)}_{l_1,m_1} a^{(1)}_{l_2,m_2} + \frac{1}{R}\Big( (l_2+1)(l_2 +2)\gamma_R + (l_2+1)(2\gamma_R + \gamma_\rho)\Big) a^{(1)}_{l_1,m_1} b^{\text{ex}(1)}_{l_2,m_2} \\
        & + \frac{(2\gamma_R + \gamma_\rho)^2}{2} a^{(1)}_{l_1,m_1} a^{(1)}_{l_2,m_2} + \frac{(l_1+1) (l_2+1)}{2R^2} b^{\text{ex}(1)}_{l_1,m_1} b^{\text{ex}(1)}_{l_2,m_2} \Bigg]H_{l,m,l_1,m_1,l_2,m_2} \\
        & + \frac{1}{2R^2}\sum_{l_1,m_1,l_2,m_2} \bigg( b^{\text{ex}(1)}_{l_1,m_1} b^{\text{ex}(1)}_{l_2,m_2}  \Theta_{l,m,l_1,m_1,l_2,m_2} -m_1m_2b^{\text{ex}(1)}_{l_1,m_1}  b^{\text{ex}(1)}_{l_2,m_2} B_{l,m,l_1,m_1,l_2,m_2} \bigg) \Bigg\},
    \end{split}
    \label{eq:second_order_eq2}
\end{equation}
\end{widetext}

To make the computation tractable, we truncate the spherical harmonic expansion at a maximum mode number of \(l_{\max}\). This truncation yields a system of $N = \sum_{l=1}^{l_{\max}}(2l+1) = (l_{\max}+1)^2-1$ distinct modes. \footnote{The \( l = 0, m = 0 \) mode is excluded from the summation because it corresponds to a perfectly spherical deformation. This effect is absorbed into the definition of \( R \) and therefore does not participate in mode coupling.} This system can then be cast into a compact matrix form:
\begin{equation}
    U\left(\dot{A}+(2\gamma_R+\gamma_\rho )A \right) + V_iB_i+ Q_i= 0,
    \label{eq:second_order_velocity_matrix}
\end{equation}
\begin{equation}
    \biggl\{\rho_i\left(U(\dot{B}_i+\ddot{R}A+gA) + P_i\right)\biggr\} \Bigg|^{i=\mathrm{ex}}_{i=\mathrm{in}} = 0.
    \label{eq:second_order_pressure_matrix}
\end{equation}
Here, $A$, $B_{\text{in}}$, and $B_{\text{ex}}$ are $N \times 1$ column vectors containing the amplitudes $a^{(2)}_{l,m}$, $b^{\text{in}(2)}_{l,m}$, and $b^{\text{ex}(2)}_{l,m}$, respectively. The elements are ordered using the index mapping
\begin{equation}
    k=l^2+l+m.
    \label{eq:k_lm_relation}
\end{equation}
The quantities $U$, $V_i$ are $N \times N$ coefficient matrices, and $P_i$, $Q_i$ are $N \times 1$ vectors, with the same index mapping. Their explicit forms are provided in Appendix~\ref{app:second_order_matrices}. Notably, the matrices $V_i$, $P_i$, and $Q_i$ are time-dependent and contain the first-order amplitudes, whereas $U$ is a constant matrix.

Eliminate $\dot{A}$ from Eq.~\eqref{eq:second_order_velocity_matrix}, we obtain
\begin{equation}
    B_\text{ex}= V_\text{ex}^{-1}(V_\text{in}B_\text{in}+Q_\text{in}- Q_\text{ex}).
    \label{eq:Bex_by_Bin}
\end{equation}
Plugging Eq.~\eqref{eq:Bex_by_Bin} into Eq.~\eqref{eq:second_order_pressure_matrix} yields a closed form evolution equation for \(B_\text{in}\):
\begin{equation}
    \begin{split}
        &(\rho_\text{in}U-\rho_\text{ex}U V_\text{ex}^{-1}V_\text{in})\dot{B}_\text{in}-\rho_\text{ex}U V_\text{ex}^{-1}\dot{V}_\text{in}B_\text{in}+ \rho_\text{in}P_\text{in}\\
        & -\rho_\text{ex}P_\text{ex}+ (\rho_\text{in}-\rho_\text{ex})(\ddot{R}+g)UA-\rho_\text{ex}U V_\text{ex}^{-1}(\dot{Q}_\text{in}- \dot{Q}_\text{ex})\\
        &-\rho_\text{ex}U \dot{V}_\text{ex}^{-1} (V_\text{in}B_\text{in}+Q_\text{in}- Q_\text{ex})=0.
    \end{split}
    \label{eq:Bin_full_evolution}
\end{equation}
We integrate this system with initial conditions \(A(t_0)=B_\text{in}(t_0)=0\) and \(\dot A(t_0)=\dot B_\text{in}(t_0)=0\), which correspond to zero initial second-order perturbations. The evolution of \(A(t)\) and \(B_\text{in}(t)\) follows Eq.~\eqref{eq:second_order_velocity_matrix} and \eqref{eq:Bin_full_evolution}.

\section{Background Setup} \label{sec:numerical}

Before integrating the system, we must prescribe the time dependence of \(R(t)\), \(\rho^i(t)\) and $\gamma_\rho(t)$. In realistic settings these arise from detailed simulations or experiments, but to explore general behavior we adopt an analytic \(R(t)\) to mimic implosion, impose the relation \(\rho^i R^3 = \mathrm{const}\) which implies  $\gamma_\rho=-3\frac{\dot{R}}{R}$, and give the initial condition $\rho^\mathrm{ex} = 5\rho^\mathrm{in}$, with Atwood number corresponds to $\alpha=-\frac{2}{3}$. We note here that the same formalism would apply equally well to an explosion or Atwood number as an arbitrary function of time.

\begin{figure}[htbp]
  \centering
  \includegraphics[width=1\linewidth]{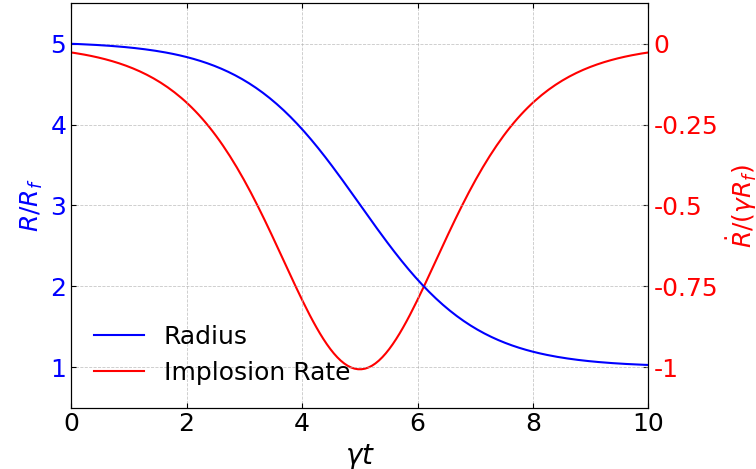}
  \caption{Demonstration of the implosion prescription: the blue curve shows \(R/R_f\) as a function of \ \(\gamma t\) (left axis), while the red curve shows \(\dot R/(\gamma R_f)\) vs.\ \(\gamma t\) (right axis).}
  \label{fig:artificial_rt}
\end{figure}

A realistic implosion rate tends to rise, then decline toward stagnation. We emulate such implosion process by constructing
\begin{equation}
    R(t)= R_f + (R_0-R_f)\frac{1+e^{-\lambda_0}}{1+e^{\gamma t-\lambda_0}},
    \label{eq:implosion_presciption}
\end{equation}
where \(R_0\) and \(R_f\) are the initial and final radii, \(\lambda_0\) controls the start stage, and \(\gamma\) controls the implosion rate (not to be confused with growth rate \(\gamma_l\)).

It should be noted that the present formula is only a simplified implosion process. For application to a more realistic system, this expression can be replaced by a more general expression, whether obtained from theoretical analysis, numerical simulations, or experimental observations.

For our numerical studies we choose \(R_0 = 5\,R_f\), \(\lambda_0 = 5\). Figure~\ref{fig:artificial_rt} shows the resulting interface radius \(R(t)\) (normalized to \(R_f\)) and implosion rate \(\dot R(t)\) (normalized to \(\gamma R_f\)) from \(\gamma t = 0\) to \(\gamma t = 10\).

\begin{figure}[htbp]
  \includegraphics[width=\linewidth]{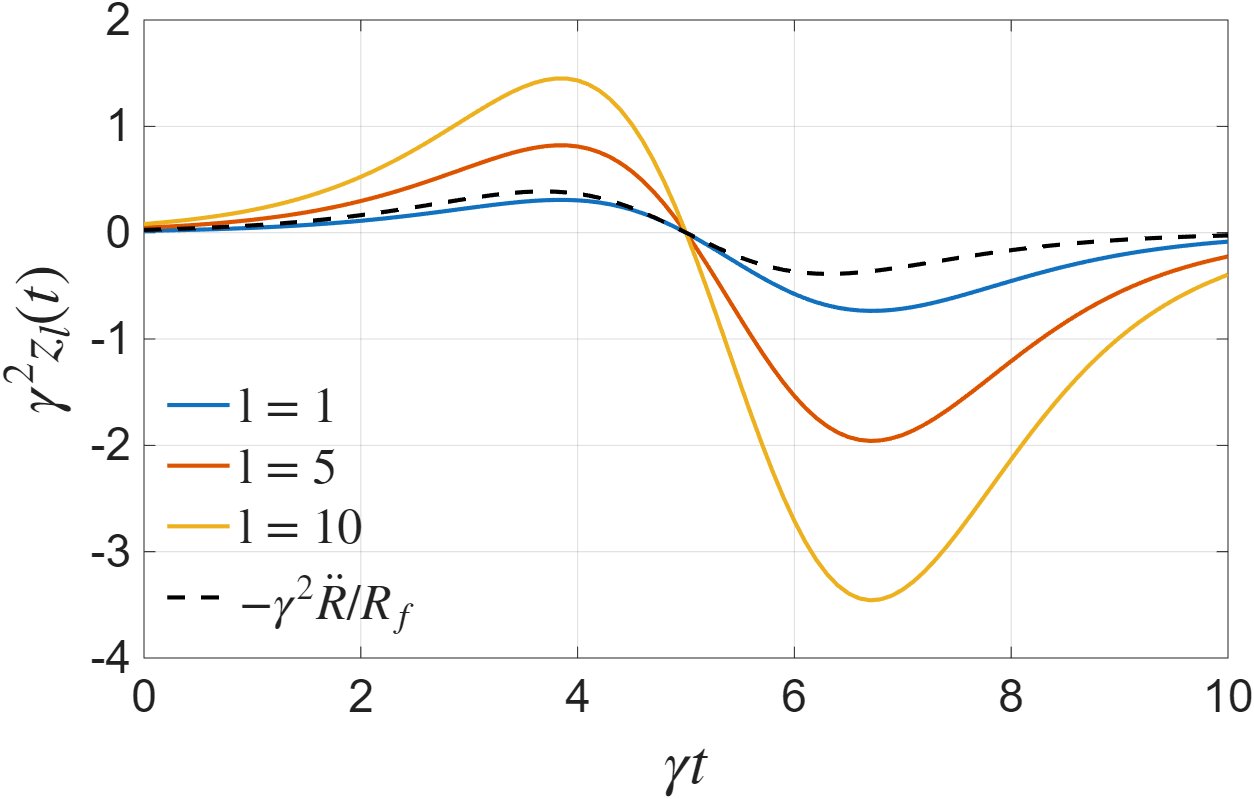}
  \caption{Time evolution of \(\gamma^2 z_l(t)\) for modes \(l = 1, 5, 10\) (solid lines), and the scaled acceleration \(-\gamma^2\,\ddot R / R_f\) (dashed line). The similar shapes highlight the tight relation between \(z_l\) and the interface acceleration.}
  \label{fig:first_order_zl}
\end{figure}

In Figure~\ref{fig:first_order_zl} we plot \(\gamma^2 z_l(t)\) for \(l = 1,5,10\) alongside the scaled acceleration \(-\gamma^2 \ddot R / R_f\) (recall from Eq.~\eqref{eq:first_order_oversimplified} that the normalized amplitude \(p_l\) satisfies $\ddot p + z_l(t)\,p = 0$).
Notably, the temporal profiles align closely, showing how the instantaneous interface acceleration directly influences \(z_l\). As \(l\) increases, the amplitude of \(\gamma^2 z_l\) also increases, consistent with the built-in scaling of \(z_l\).  When \(\ddot R = 0\), \(\gamma^2 z_l\) crosses zero (unlike static sphere cases where gravitational acceleration enforces a nonzero value). Early in the implosion, \(\gamma^2 z_l\) is positive, implying oscillatory behavior. However, during the deceleration stage it becomes negative, driving exponential-like growth. This suggests that in a converging interface, higher-\(l\) modes experience more intense early oscillations and more rapid late growth, in qualitative alignment with static background expectations.

During the early stage of the implosion (\(\gamma t < 5\)), the interface acceleration is directed outward, leading to \(z_l < 0\) and suppressing the Rayleigh–Taylor instability (RTI). In this regime, the perturbation amplitudes merely oscillate rather than grow, rendering it unimportant for studying RTI. Therefore, we initiate the perturbation evolution at \(\gamma t = 5\).

For the full evolution, including both first- and second order amplitudes, the physical requirement that the interface perturbation \(\eta(\theta,\varphi,t)\) must be real-valued imposes the condition
\[
a_{l,-m}(t) = (-1)^m\,a^{*}_{l,m}(t)\;.
\]
Thus, initializing a single complex spherical harmonic mode would yield a non-physical complex interface shape. To ensure real-valued perturbations, we consider single-mode static initial conditions with real amplitudes and vanishing initial velocities at \(\gamma t_0 = 5\):
\[
a_{l_0,m_0}^{(1)}(t_0) = (-1)^{m_0} a^{(1)}_{l_0,-m_0}(t_0)=\,\eta_0, \quad  \dot{a}^{(1)}_{l,m} = 0,
\]
for the representative cases \((l_0,m_0) = (3,1),\,(4,0),\,(4,1),\,(4,4)\). The axisymmetric mode (\(m=0\)) is inherently real, so we set \(a_{4,0}(t_0) = 2\,\eta_0\) to ensure same amplitude. In all cases, we take \(\eta_0/R_f = 0.002\).

To explore more physically realistic RTI behavior, we further initialize the interface with a Gaussian-shaped bubble centered on the north pole:
\begin{equation}
\eta(\theta, \varphi, t_0) = 0.003 R_f \times \exp(-10\theta^2).
\label{eq:bubble_init_condition}
\end{equation}
This configuration mimics a localized deformation of the interface and illustrates how such a perturbation evolves over time.

As discussed in Section~\ref{sec:equations}, we truncate the mode at $l_\text{max}=16$ to avoid unaffordable computational cost. For most interested low-$l$ modes, $l_\text{max}=16$ ensures acceptable numerical convergence (details can be found in Appendix~\ref{app:lmax_convergence}).

\section{Numerical Results} \label{sec:results_discussions}

In this section, we present the outcomes of the numerical integration of the first and second order perturbation equations and interpret their physical consequences. We begin with the single mode evolution, highlighting how a dynamic background (i.e., Bell–Plesset effects) modifies first order behavior relative to static spherical geometry. We then turn to multi-mode interactions arising from second order coupling, exploring energy diffusion among modes and its impact on interface growth. After that, we analyze the ``bubble"-like structure evolution, studying how multi-mode initial condtion influence our results. Finally, we analyze the influence of the BP effects, highlighting its role in amplifying the instability.

\subsection{First Order Evolution} \label{subsec:first_order_results}

\begin{figure}[htbp]
  \centering
  \includegraphics[width=1\linewidth]{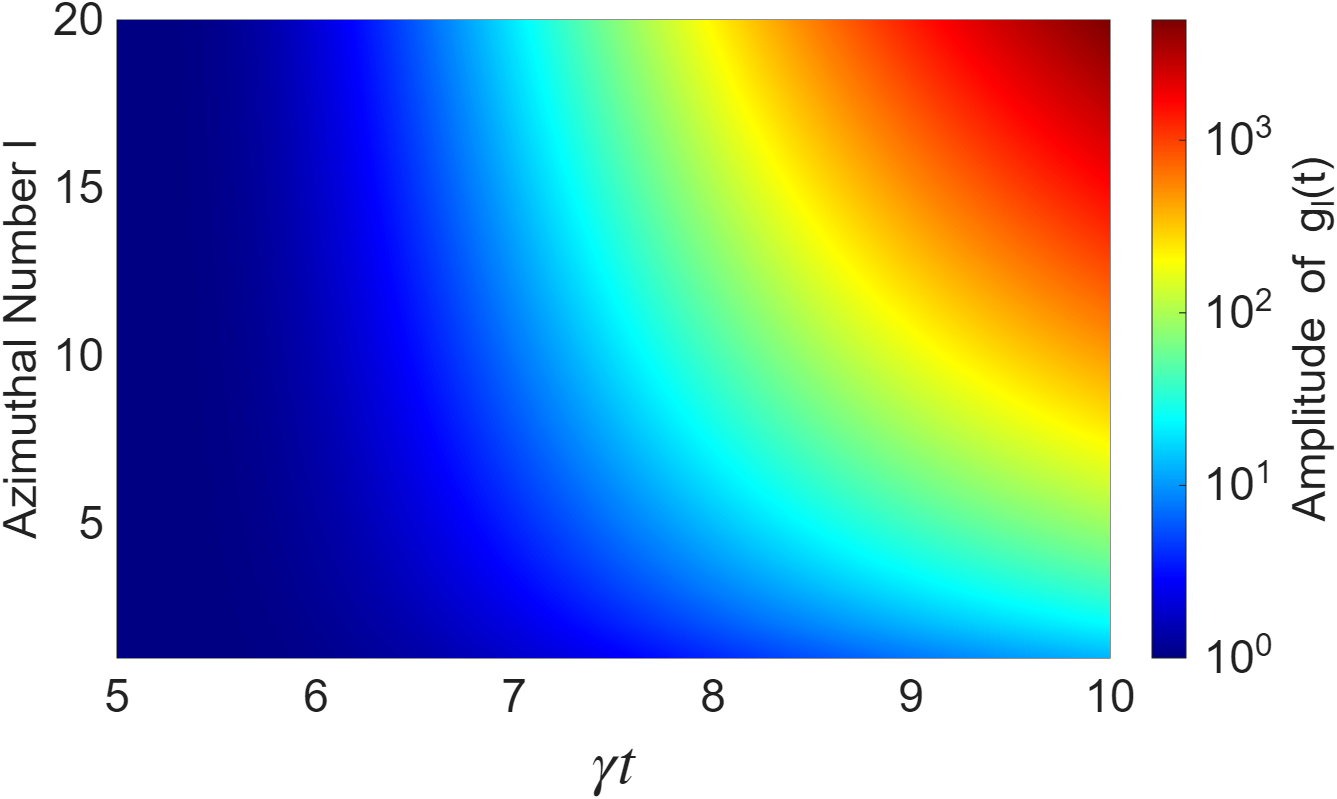}
  \caption{Heatmap of \(g_l(t)\) for \(l = 1\) to \(20\) evolving from $\gamma t=5$ to $10$. Redder regions denote large amplitudes, while blue regions denote small amplitudes.}
  \label{fig:first_order_gl_heatmap}
\end{figure}

At the first order, the different spherical harmonic modes evolve independently. Figure~\ref{fig:first_order_gl_heatmap} presents a heatmap of the first order growth factor, \(g_l(t)\), for poloidal mode numbers ranging from \(l=1\) to \(20\). The visualization clearly illustrates a primary characteristic of the first order amplitudes that higher \(l\) modes grow faster over the same time interval, consistent with the larger $\gamma_l$ (Eq.~\eqref{eq:reduced_growth_factor}) found in static background.

\begin{figure}[htbp]
  \centering
  \includegraphics[width=0.9\linewidth]{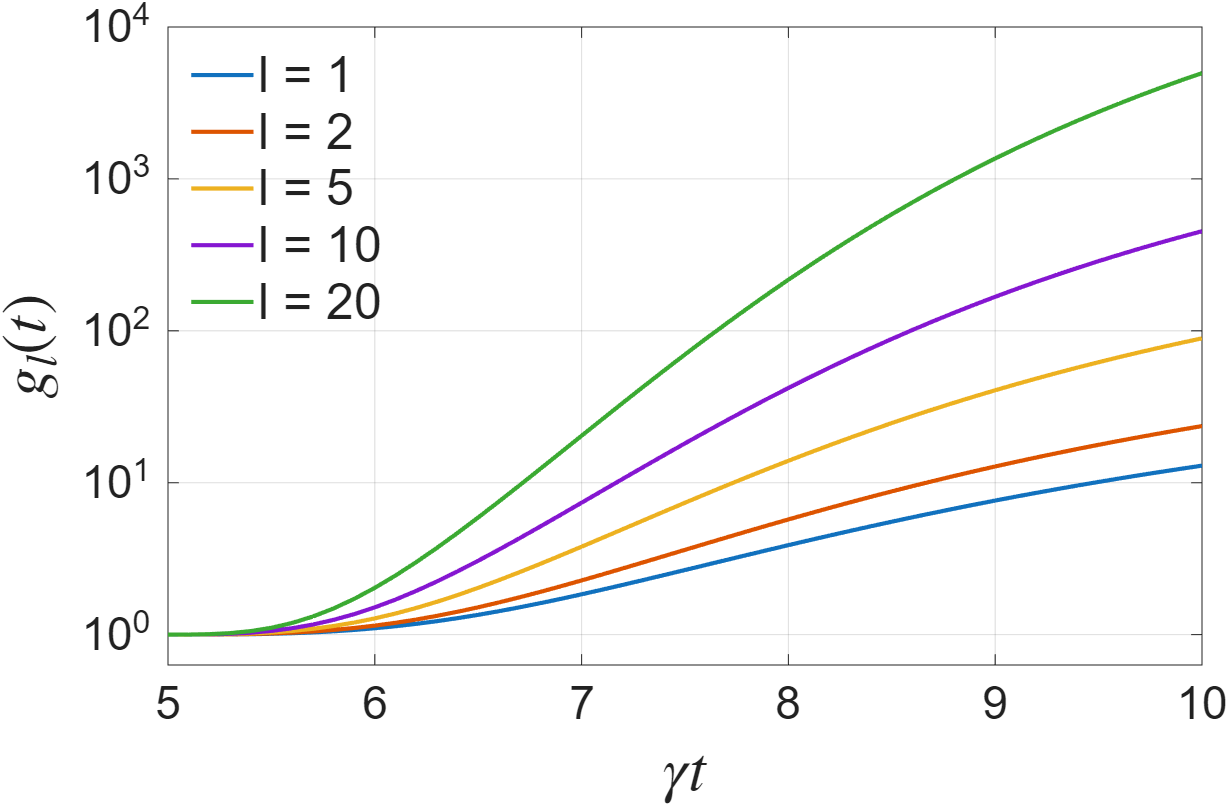}
  \caption{Evolution of the growth factor \(g_l(t)\) for selected modes (\(l = 1, 2, 5, 10, 20\)) plotted on a logarithmic scale. The curves illustrate a period of accelerated growth followed by a saturation into a linear growth regime at late times as the interface deceleration ceases.}
  \label{fig:first_order_gl_plot}
\end{figure}

To illustrate individual mode behavior, Figure~\ref{fig:first_order_gl_plot} plots \(g_l(t)\) for selected \(l=1,2,5,10,20\) in log scale. The amplitudes start with an initial period of slow amplification, followed by a phase of rapid, super-exponential growth, and finally a transition toward slower, linear growth as the driving term \(z_l(t) \to0\) at last in our prescribed implosion scenario. While higher-\(l\) modes achieve much larger amplitudes, the temporal profile of their growth is qualitatively similar across all modes, in correspondance with the $z_l$ shown in Figure~\ref{fig:first_order_zl}.
Our results show that in a dynamically imploding background, the first order evolution behaves qualitatively similar to static sphere models, though BP effects enhances the instability growth by adding extra terms in Eq.~\eqref{eq:first_order_eq}.

\subsection{Second Order Evolution} \label{subsec:second_order_results}

Moving to the second order analysis, we now solve the full coupled system to investigate the nonlinear dynamics, with the four different initial conditions \((l,m) = (3,\pm1),\,(4,0),\,(4,\pm1),\,(4,\pm4)\).

\begin{figure}[htbp]
  \centering
  \includegraphics[width=1\linewidth]{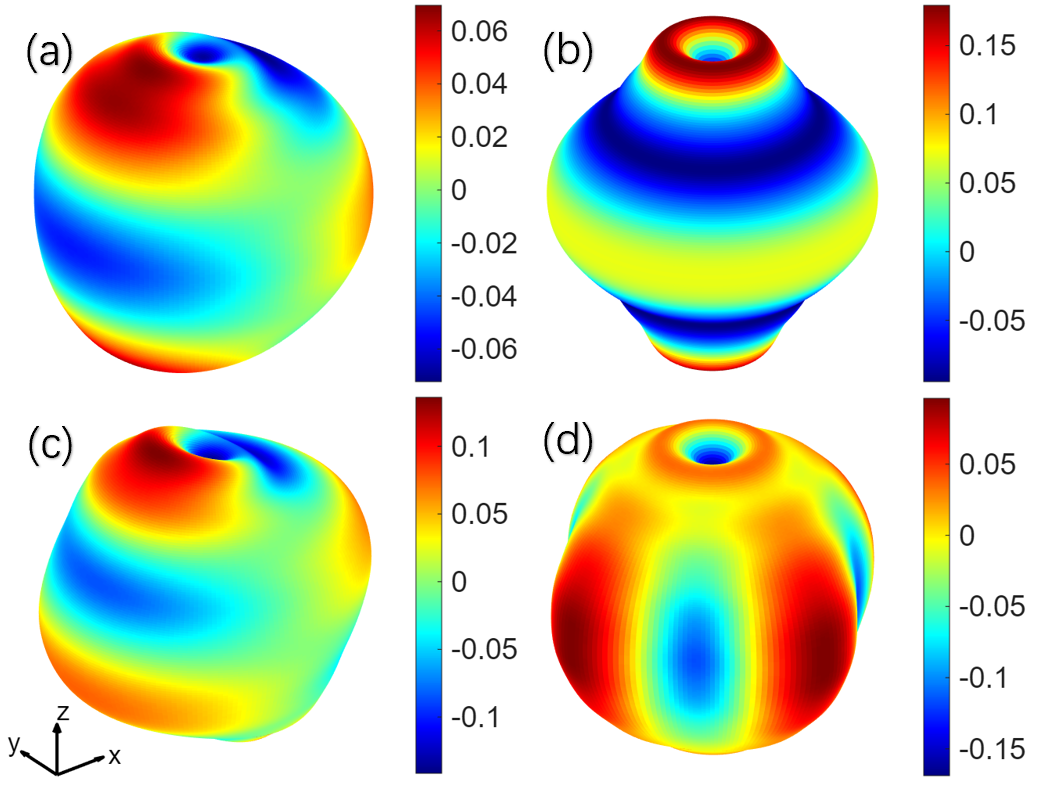}
  \caption{Interface shape at \(\gamma t = 10\) for four initial conditions: \((l_0,m_0) = \text{(a) }(3,\pm1),\,\text{(b) }(4,0),\,\text{(c) }(4,\pm1),\,\text{(d) }(4,\pm4)\). Colors represent the normalized displacement amplitude ($\eta/R_f$), where red indicates outward displacement (\(\eta > 0\)) and blue indicates inward displacement (\(\eta < 0\)). The directions of coordinates are marked by the arrows on the bottom left.}
  \label{fig:surface_all}
\end{figure}

Figure~\ref{fig:surface_all} displays the interface morphology for each case at the final time \(\gamma t = 10\). The surfaces exhibit complex, multi-modal topographies, a stark departure from the simple, single-mode structure of the initial conditions. This complexity is a direct consequence of nonlinear mode coupling, which redistributes energy from the initial modes to a broad spectrum of other modes.

A striking feature in Figure~\ref{fig:surface_all} is the markedly larger deformation observed for the axisymmetric \((l,m)=(4,0)\) case compared to the others. This observation strongly suggests that axisymmetric perturbations are predisposed to more vigorous growth. The spectral analysis presented next provides a quantitative confirmation of this behavior.

\begin{figure*}[htbp]
  \centering
  \includegraphics[width=\linewidth]{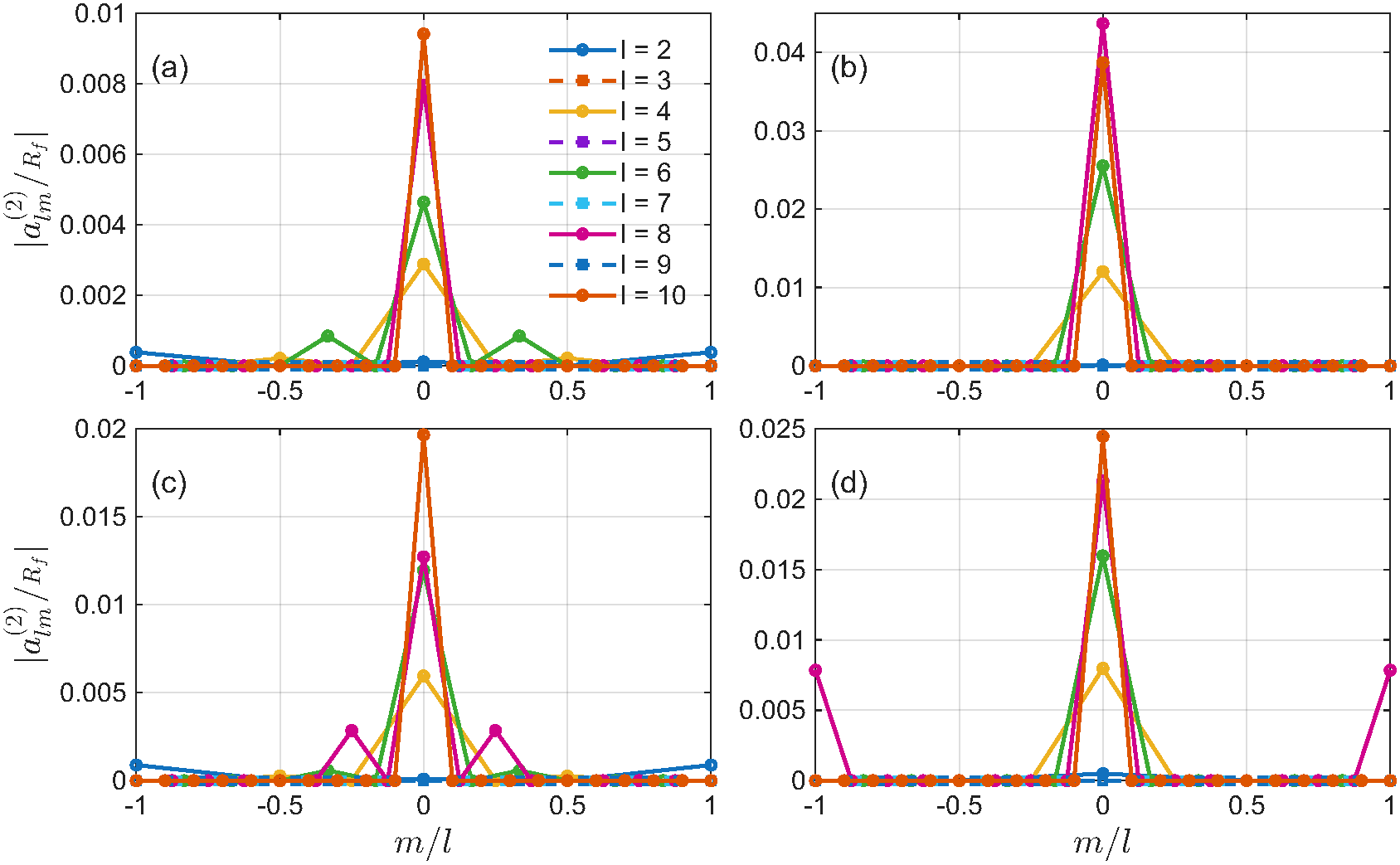}
  \caption{Spectra of normalized second-order amplitudes \(|a^{(2)}_{l,m}|/R_f\) at \(\gamma t = 10\) for four initial conditions: \((l_0,m_0) = \text{(a) }(3,\pm1),\,\text{(b) }(4,0),\,\text{(c) }(4,\pm1),\,\text{(d) }(4,\pm4)\). The horizontal axis is the normalized azimuthal index \(m/l\). Solid curves (even \(l\)) and dashed curves (odd \(l\)) show a powerful selection rule: only the equatorially symmetric (even \(l\)) modes are excited, and among these, the axisymmetric (\(m=0\)) modes are dominant.}
  \label{fig:spectrum_all}
\end{figure*}

To quantify this mode redistribution, Figure~\ref{fig:spectrum_all} shows the spectral distribution of the second-order amplitude magnitudes, \(|a^{(2)}_{l,m}|\), at \(\gamma t = 10\). As required by the real-valued nature of the perturbation, the time evolution preserves the symmetry $a_{l,-m}(t) = (-1)^m a^*_{l,m}(t)$, resulting in a spectrum that is symmetric in \(m\), up to negligible numerical error. The spectra reveal two powerful selection rules. First, for all initial conditions, energy is transferred exclusively to even-\(l\) modes, all odd-\(l\) amplitudes at second order remain identically zero. Second, among the excited even-\(l\) modes, nonlinear interactions preferentially channel energy into the axisymmetric (\(m=0\)) component.

This tendency should be understood with respect to the intrinsic symmetry axis of the evolving perturbation. In our examples, the initial conditions single out a natural axis (e.g., the $z$-axis for the $(4,0)$ case), and in a spherical-harmonic basis aligned with that axis, an axisymmetric structure is represented entirely (or predominantly) by \(m=0\) modes; in particular, for the purely axisymmetric $(4,0)$ initial condition, the evolution remains axisymmetric and therefore excites only \(m=0\) modes up to second order. If instead one chooses a coordinate system whose \(z\)-axis is not aligned with the perturbation’s symmetry axis, the same axisymmetric structure is decomposed into a broad distribution of \(m\neq 0\) components. This coordinate dependence, and the corresponding rotation invariance of the physical evolution, will be demonstrated explicitly in Section~\ref{subsec:bubble_results}.

To explain this phenomenon, we must examine the structure of the coupling coefficients in Eqs.~\eqref{eq:second_order_eq1_in}-\eqref{eq:second_order_eq2}. The $B_{l,m,l_1,m_1,l_2,m_2}$, $H_{l,m,l_1,m_1,l_2,m_2}$, and $\Theta_{l,m,l_1,m_1,l_2,m_2}$ terms act as the source terms that couple two first-order quantities to generate a second-order amplitude. The $A_{l,m,l_1,m_1}$ term governs the subsequent couplings among the second-order amplitudes themselves. As detailed in Appendix~\ref{app:second_order_equation}, all these coefficients are derived from integrals over products of several spherical harmonics.

The key to the selection rule lies in the inversion symmetry (parity) of spherical harmonics:
\begin{equation}
Y_l^m(\pi-\theta,\pi+\varphi) = (-1)^l Y_l^m(\theta,\varphi).
\label{eq:spherical_harmonics_parity}
\end{equation}
For an integral over the full sphere of a product of spherical harmonics ($\prod_i Y_{l_i}^{m_i}$) to be non-zero, the integrand must be an even function under this transformation, as the integral of any odd function over this symmetric domain is zero. This imposes a strict parity selection rule: the integral vanishes unless the sum of the polar indices, $\sum l_i$, is an even number.

We first apply this rule to the source terms ($B$, $H$, $\Theta$). These coefficients couple two first-order modes ($l_1$, $l_2$) to generate a second-order mode ($l$). The integral is non-zero only if $l_1 + l_2 + l = \text{even}$. In our case, the first-order modes originate from the same initial mode, $l_0$, so $l_1 = l_2 = l_0$. The condition for excitation thus becomes $l_0 + l_0 + l = \text{even}$. Since $l_0 + l_0$ is always even, this rigorously forces the resulting second-order mode $l$ to be even.

Once these even-$l$ modes are generated, the $A_{l,m,l_1,m_1}$ coefficient, which governs their mutual interactions, must also obey the same parity rule. This ensures that an even $l$ mode can only couple to another even $l_1$ mode, creating a closed system for even-$l$ modes. As a result, energy diffuses exclusively among even-$l$ modes. This is precisely what is observed in Figure~\ref{fig:spectrum_all}: only the even-$l$ amplitudes (solid lines) are excited, while all odd-$l$ amplitudes (dashed lines) are zero.

\begin{figure*}[htbp]
  \centering
  \includegraphics[width=1\linewidth]{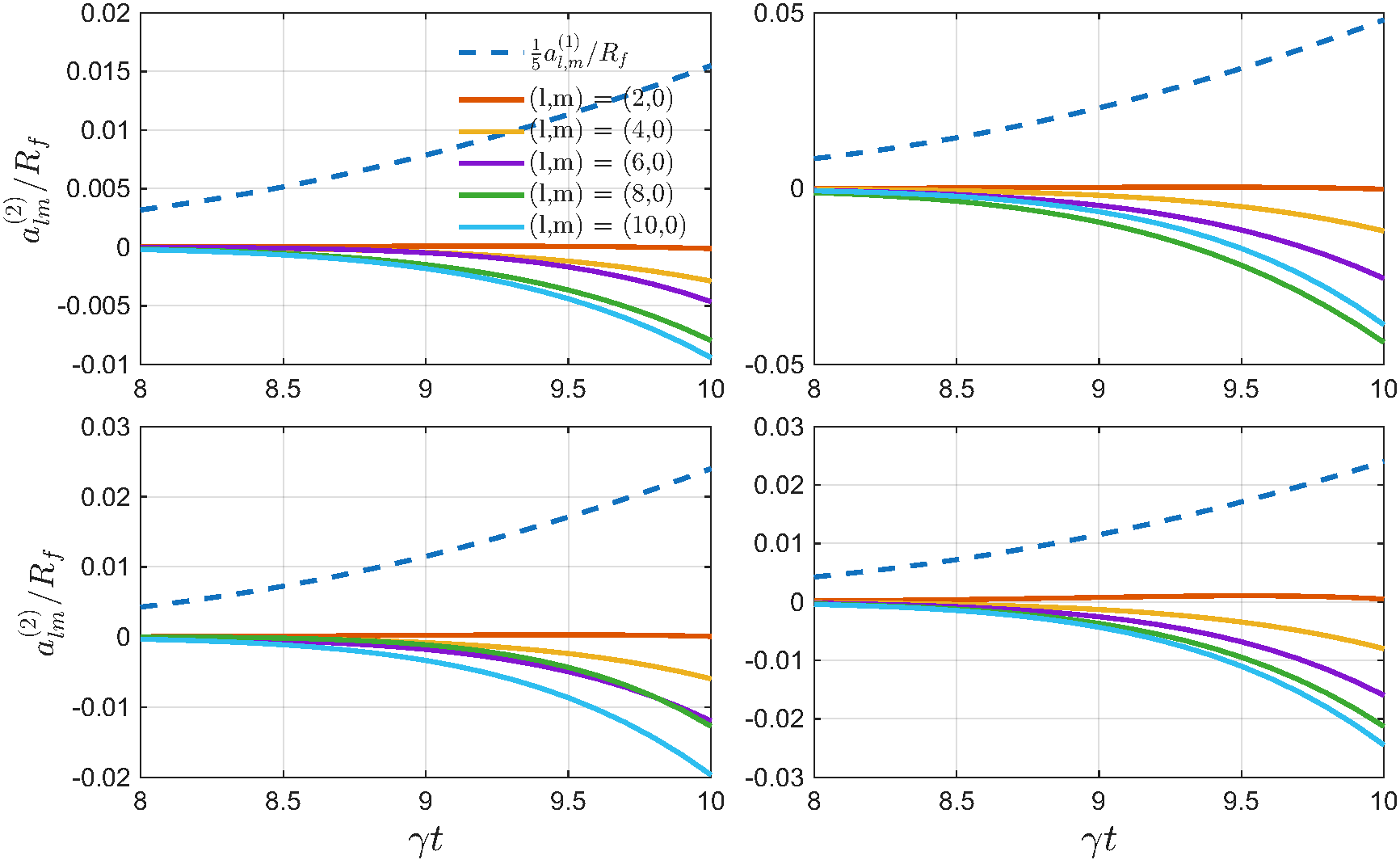}
  \caption{Evolution of the first order amplitude (dashed line) and the dominant second order amplitudes (solid lines, scaled by 1/5 for visibility) for the four initial conditions: \((l_0,m_0) = \text{(a) }(3,\pm1),\,\text{(b) }(4,0),\,\text{(c) }(4,\pm1),\,\text{(d) }(4,\pm4)\). The second order amplitudes show complex growth but remain significantly smaller than the first order mode.}
  \label{fig:evolution_all}
\end{figure*}

Additionally, a second selection rule must be satisfied for the azimuthal index: $\sum m_i=0$. This arises because any un-cancelled factor $e^{i\varphi\sum m_i}$ in the integrand integrates to zero over $\varphi \in [0, 2\pi]$. The $m=0$ modes are unique in several aspects. For a purely axisymmetric initial condition (containing only $m=0$ modes), energy transfers exclusively to $m=0$ modes up to second order. More generally, they can be generated by any self-coupling ($m_1 = m_0, m_2 = -m_0$) and participate in the largest number of allowed coupling pathways. This ``coupling promiscuity" likely explains their preferential growth and observed dominance in the spectra. This analysis validates our finding that the axisymmetric ($m=0$) modes are the fastest growing.

Notably, our results indicate that the azimuthal mode number $m$ of the initial perturbation plays a crucial role in determining the pattern of the instability, whereas the poloidal mode number $l$ primarily governs the instability growth rate. For example, the interface morphology and spectrum for $(l,m)=(3,\pm 1)$ (Figure~\ref{fig:surface_all} (a), \ref{fig:spectrum_all} (a)) bear a qualitative resemblance to those of the $(l,m)=(4,\pm 1)$ case (Figure~\ref{fig:surface_all} (c), \ref{fig:spectrum_all} (c)). Moreover, the three cases with $l=4$ but different $m$ values, namely $(4,0)$, $(4,\pm 1)$, and $(4,\pm 4)$, yield distinctly different outcomes. In contrast, the perturbation growth rates for the three cases with the same $l$ value are comparable, and significantly larger than that of the $l=3$ case.

\begin{figure}[htbp]
  \centering
  \includegraphics[width=1\linewidth]{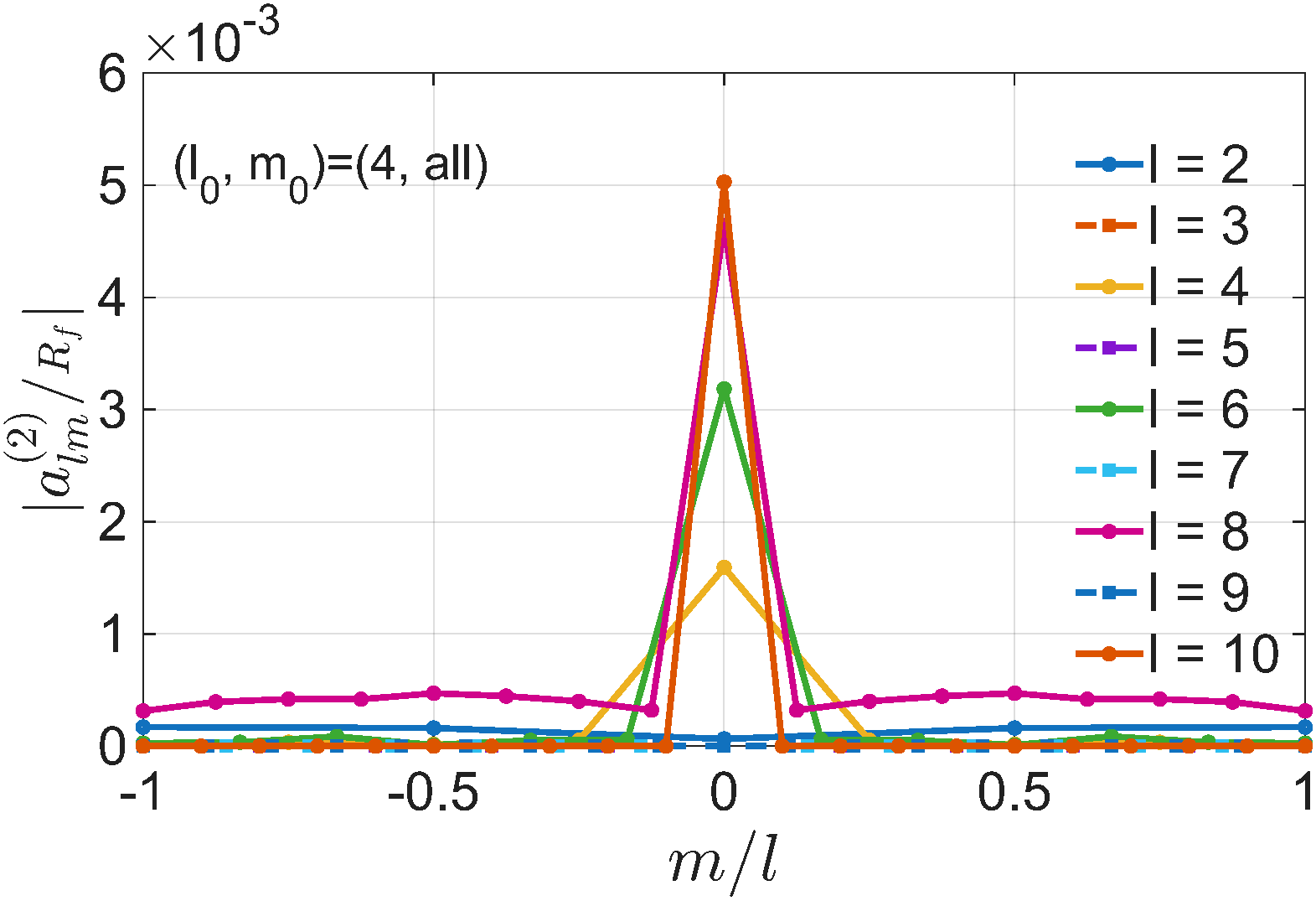}
  \caption{Spectrum of normalized second order amplitudes \(\bigl|a^{(2)}_{l,m}\bigr|/R_f\) for \(l = 2\) to 10 at \(\gamma t = 10\), for the initial condition $a^{(1)}_{4,m_0\ne0}(t_0)=\frac{(-1)^{m_0}}{5}\eta_0$ and $a^{(1)}_{4,0}(t_0)=\frac{2}{5}\eta_0$.}
  \label{fig:spectrum_l4m_all}
\end{figure}

To compare the growth of the orders, Figure~\ref{fig:evolution_all} illustrates the time evolution of the dominant second order modes ($l=2,6,8,12$, $m=0$) alongside the initial first order amplitude (scaled down by a factor of five). The second order amplitudes display accelerating growth rates that eventually surpass that of the first order. Notably, they remain significantly smaller in magnitude than the first order amplitude throughout the simulation, while the first order amplitude itself remains much smaller than the unperturbed interface radius. This result is crucial, as it validates the applicability of our perturbative method up to $\gamma t=10$, confirming that the system remains within the weakly nonlinear regime and our model provides an accurate assessment of the instability. Furthermore, the plots reveal that the dominant second order amplitudes tend to develop negative signs, whereas other modes like $(2,0)$ exhibit positive values, highlighting the complex phase dynamics that are not present in the first order growth. \footnote{In the single-mode examples considered here, the dominant second-order amplitudes happen to be predominantly negative. This sign coherence is not a universal feature: for more general multi-mode initial conditions (e.g., the bubble tests), both positive and negative second-order amplitudes appear; see Appendix~\ref{app:bubble_detail}.}

Additionally, to examine whether the axisymmetric tendency arises specifically from single-mode initial conditions, we initialize a multi-mode perturbation containing all $m_0$ values for $l_0=4$, with the initial amplitudes reduced by a factor of 5 to maintain comparable total perturbation energy ($a^{(1)}_{4,m_0\neq0}(t_0)=\frac{(-1)^{m_0}}{5}\eta_0$ and $a^{(1)}_{4,0}(t_0)=\frac{2}{5}\eta_0$). The final spectrum is shown in Figure~\ref{fig:spectrum_l4m_all}. Despite the larger number of excited modes, the $m=0$ amplitudes remain dominant, validating our previous observations. Further analysis of how the axisymmetric tendency varies with the initial mode number $l_0$ is presented in Appendix~\ref{app:axisymmetric_tendency}.

\subsection{"Bubble"-like Structure Evolution} \label{subsec:bubble_results}

\begin{figure}[htbp]
  \centering
  \includegraphics[width=1\linewidth]{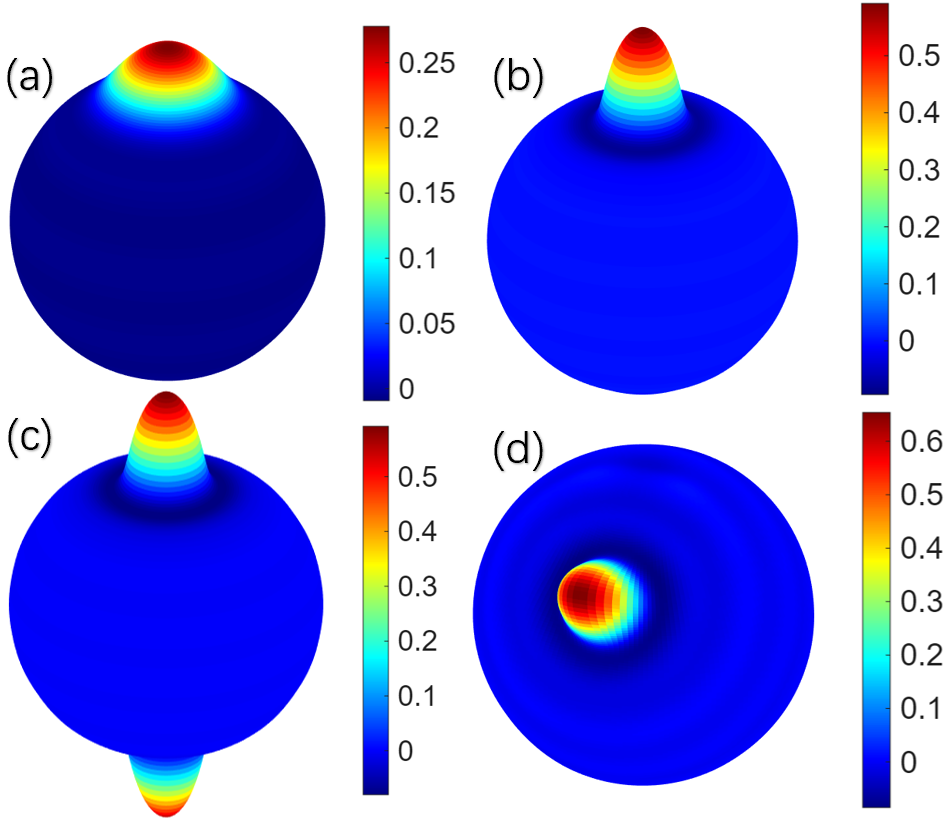}
  \caption{Interface evolution for the ``bubble" perturbations. (a) The initial perturbation shape, reconstructed from the $l \le 16$ spectral projection. The amplitudes are scaled 100x for visibility purpose. Evolved interface at \(\gamma t = 10\) for: (b) a single bubble at the north pole; (c) two identical bubbles at the north and south poles; (d) a single bubble rotated to the $(-1,-1,1)/\sqrt{3}$ direction.}
  \label{fig:surface_bubble}
\end{figure}

We examine the RTI growth from a more complex, multi-mode initial condition designed to simulate a localized ``bubble" perturbation (Eq.~\eqref{eq:bubble_init_condition}). To construct this state, we first define a desired initial surface perturbation, $\eta(\theta, \varphi, t_0)$, and then project it onto the spherical harmonic basis to obtain the initial modal amplitudes, $a^{(1)}_{l_0,m_0}(t_0)$ via
\begin{equation}
a^{(1)}_{l_0,m_0}(t_0) = \int \eta(\theta, \varphi, t_0) Y^*_{l_0,m_0} \mathrm{d}\Omega.
\label{eq:Y_projection}
\end{equation}

To examine whether the placement of the bubble alters the perturbation growth, we consider three additional initial conditions:
\begin{itemize}
    \item First, a single bubble at the north pole, as defined in Eq.~\eqref{eq:bubble_init_condition}.
    \item Second, two identical bubbles placed at the north and south poles. Due to the symmetry with respect to the equatorial plane and the selection rules discussed in Section~\ref{subsec:second_order_results}, only even-$l$ modes are excited throughout the evolution.
    \item Third, a single bubble rotated to an arbitrary direction, aligned with the unit vector $(-1,-1,1)/\sqrt{3}$.
\end{itemize}

Figure~\ref{fig:surface_bubble} shows the initial interface and the evolved interface shapes at $\gamma t = 10$ for the three cases. The initial shape, as reconstructed from our $l \leq 16$ spectral expansion, demonstrates that our truncation adequately captures the localized Gaussian perturbation. The subsequent evolution shows that the bubble grows locally, largely retaining its shape while increasing in amplitude. Notably, in the single-bubble cases, the odd-$l$ and even-$l$ modes counteract at the opposite pole, maintaining the localization of the perturbation, as demonstrated in Appendix~\ref{app:bubble_detail}. The bubble growths in the three cases remain consistent with each other despite distinctly different mode compositions, within numerical errors arising from the finite spherical harmonic truncation. These results validate that our method is independent of the perturbation location.

\begin{figure}[htbp]
  \centering
  \includegraphics[width=1\linewidth]{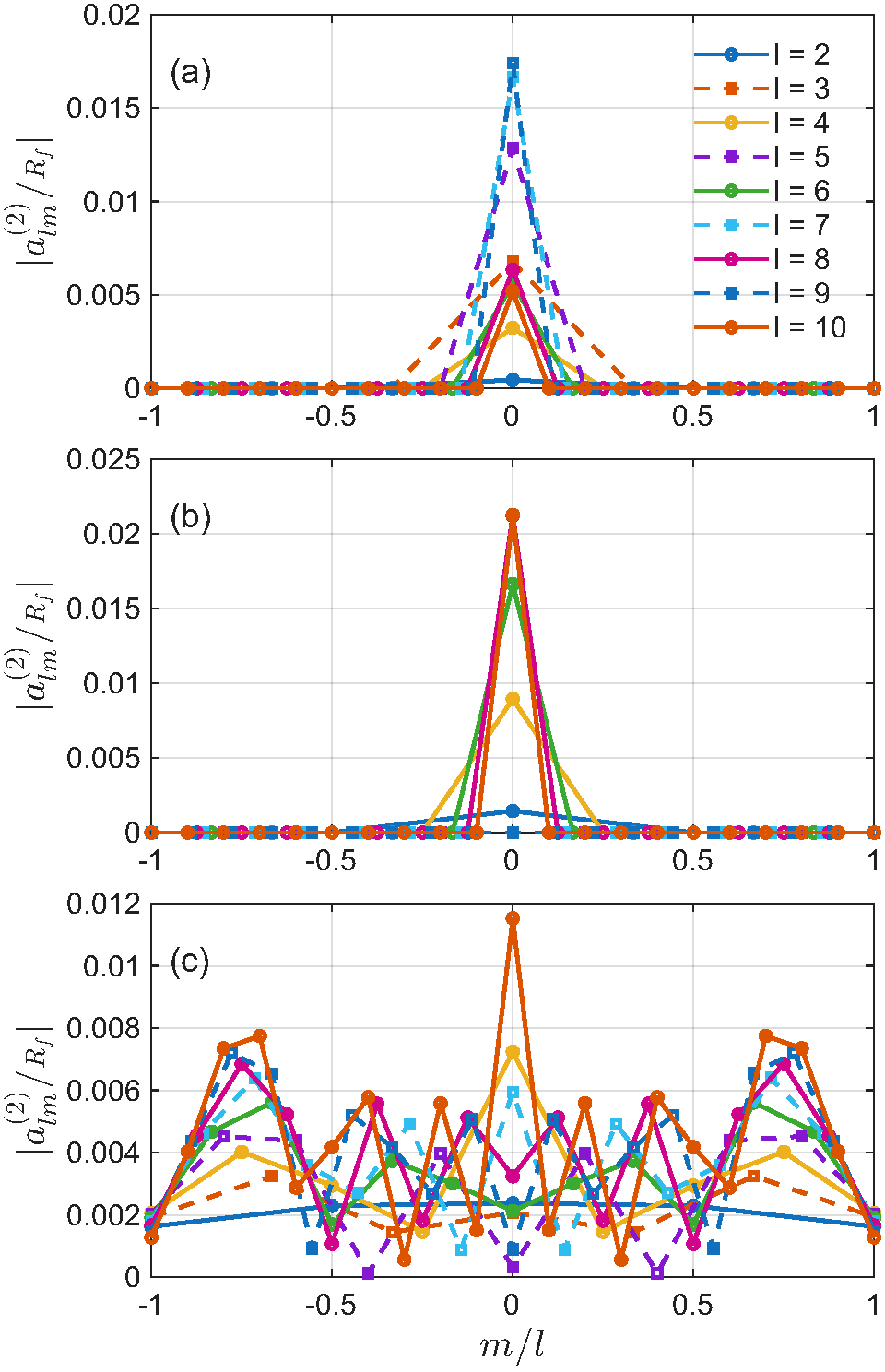}
  \caption{Spectrum of normalized second order amplitudes \(|a^{(2)}_{l,m}|/R_f\) for \(l=2\) to \(10\) at \(\gamma t = 10\), resulting from the multi-mode ``bubble" initial conditions: (a) a single bubble at the north pole; (b) two identical bubbles at the north and south poles; (c) a single bubble rotated to the $(-1,-1,1)/\sqrt{3}$ direction.} The horizontal axis is the normalized azimuthal index \(m/l\). Solid (dashed) lines represent even (odd) \(l\).
  \label{fig:spectrum_bubble}
\end{figure}

Figure~\ref{fig:spectrum_bubble} shows the resulting second-order spectral distributions for the three cases at $\gamma t=10$. Since the initial ``bubble" state contains a superposition of both even and odd $l$ modes, the selection rule $l_1+l_2+l = \text{even}$ now permits the excitation of odd-$l$ second-order amplitudes, which are clearly visible in the spectrum. Moreover, the amplitudes alternate between positive and negative values, as shown in Appendix~\ref{app:bubble_detail}. Despite the excitation of this broader range of modes, the axisymmetric ($m=0$) modes still exhibit a tendency to dominate the nonlinear growth. For the double-bubble case, no odd-$l$ modes appear, as argued above; however, the even-$l$ modes behave similarly to the single-bubble case but with larger amplitudes. For the rotated bubble, the spectrum exhibits a more complex structure, and the axisymmetric tendency appears weaker in the chosen coordinate system. However, this is merely a consequence of projecting an axisymmetric perturbation onto a misaligned coordinate frame. In a coordinate system with its $z$-axis aligned with the bubble's symmetry axis, the $m=0$ modes would remain dominant, as in the north-pole case. These results reinforce our previous conclusion regarding the robust and preferential growth of axisymmetric perturbations, while also demonstrating that this physical tendency is independent of the arbitrary choice of coordinate axes.

\begin{figure}[htbp]
  \centering
  \includegraphics[width=1\linewidth]{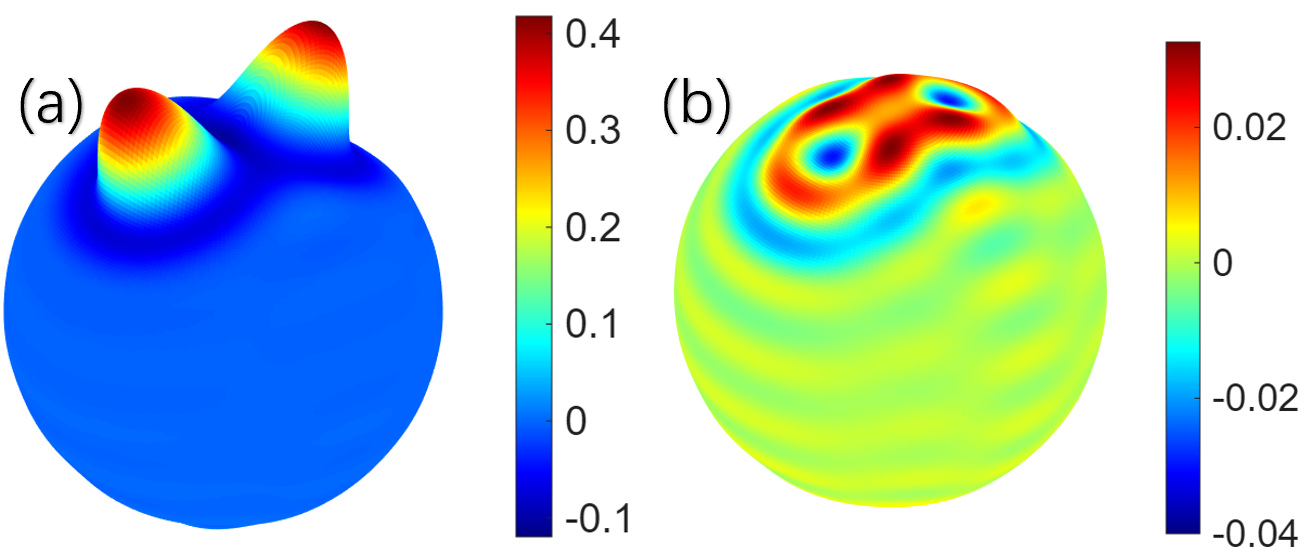}
  \caption{Interface evolution for the twin bubbles centered along the directions $(\pm 5/12,\,0,\,5/13)$: (a) full perturbation including both first- and second-order amplitudes; (b) contribution from second-order amplitudes only.}
  \label{fig:surface_bubble_twins_details}
\end{figure}

\begin{figure}[htbp]
  \centering
  \includegraphics[width=1\linewidth]{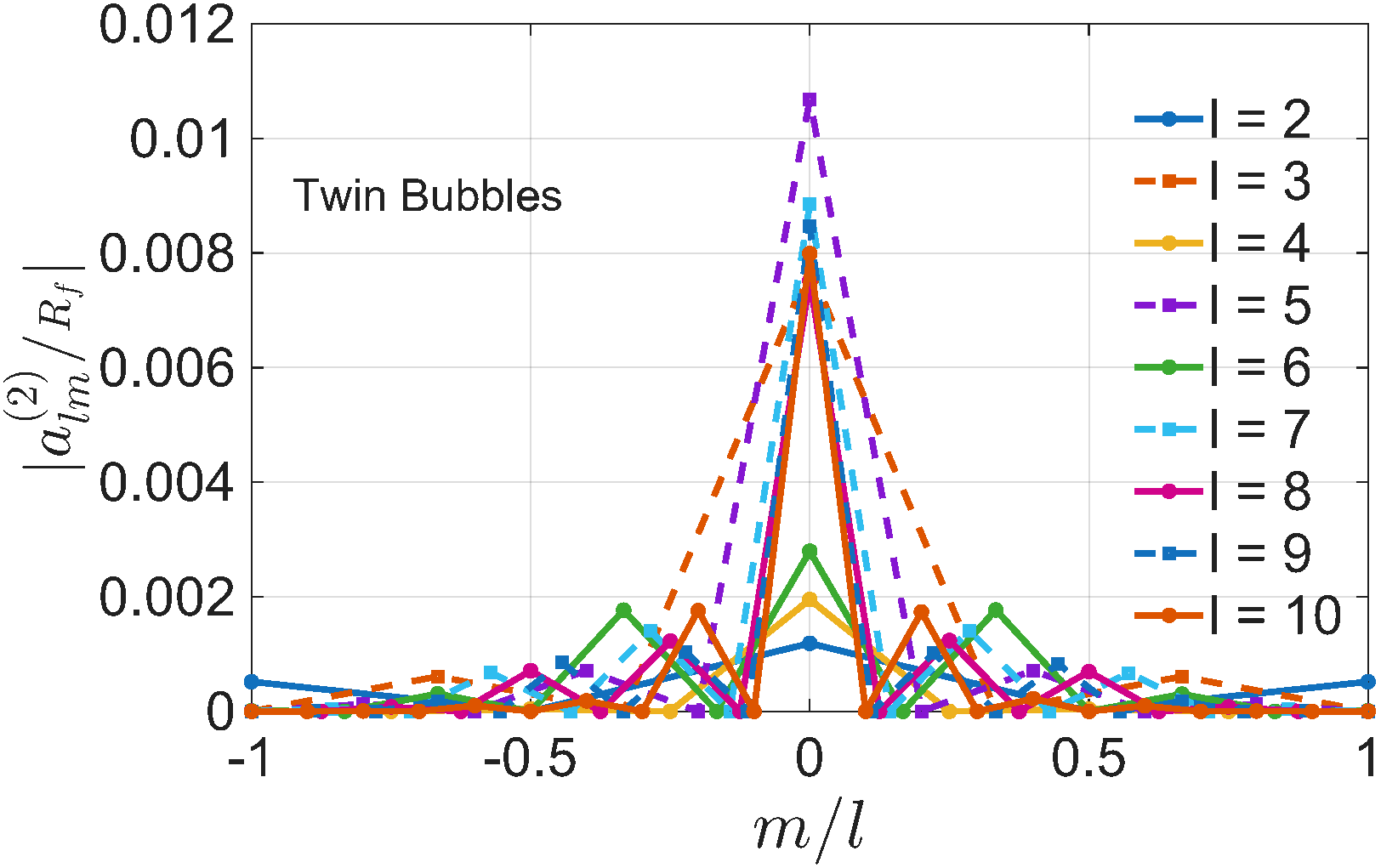}
  \caption{Spectrum of normalized second order amplitudes \(|a^{(2)}_{l,m}|/R_f\) for \(l=2\) to \(10\), resulting from the twin bubbles centered along the directions $(\pm 5/12,\,0,\,5/13)$. Solid (dashed) lines represent even (odd) \(l\).}
  \label{fig:spectrum_bubble_twins}
\end{figure}

Additionally, we perform a simulation with twin bubbles initialized using identical profiles but with centers in close proximity, oriented along the directions $(\pm 5/13,\,0,\,12/13)$, to investigate how neighboring bubbles interact. Figure~\ref{fig:surface_bubble_twins_details}(a) and Figure~\ref{fig:spectrum_bubble_twins} show the evolved interface and spectrum, respectively. Although the second-order amplitudes are comparable to those of the three cases discussed above, the evolved bubbles are noticeably smaller. At first order, all bubbles should grow independently and reach roughly the same size, so this difference must arise from higher-order effects. To clarify this, Figure~\ref{fig:surface_bubble_twins_details}(b) isolates the second-order contribution to the interface by excluding the first-order amplitudes. Rather than amplifying the first-order growth, the second-order amplitudes partially cancel one another, slightly shrinking the bubbles. This suggests that two closely spaced bubbles interact at second order in a manner that suppresses the overall perturbation growth, rather than promoting coalescence.

\begin{figure}[htbp]
  \centering
  \includegraphics[width=0.9\linewidth]{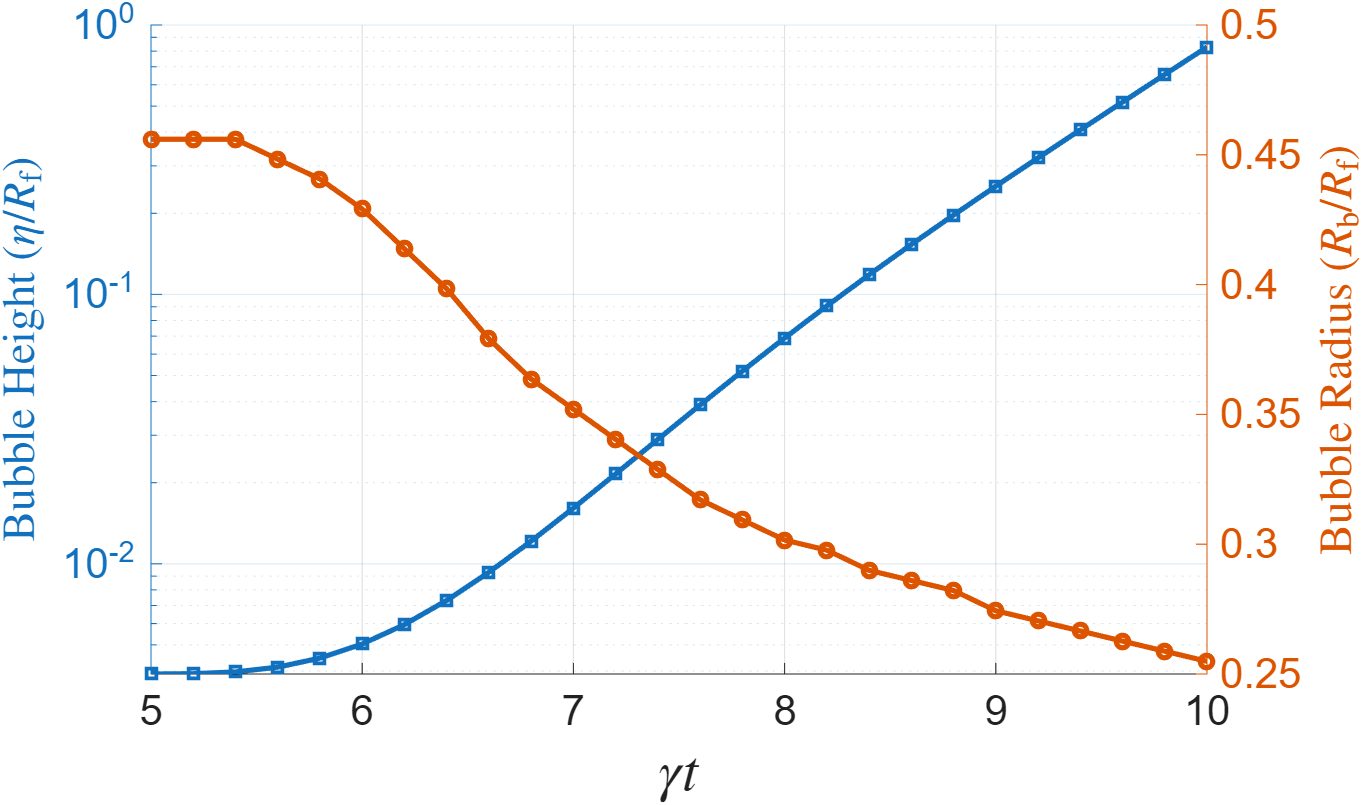}
  \caption{Evolution of the normalized bubble height ($\eta_{\text{peak}}/R_f$, blue-square curve) and bubble radius ($R_b/R_f$, orange-circle curve) for the multi-mode ``bubble" case. The radius $R_b$ is defined as the location where the amplitude drops to 1/10 of the peak.}
  \label{fig:bubble_growth}
\end{figure}

Finally, to better quantify the instability growth, we track two metrics for the single bubble at the north pole: its peak height ($\eta_{\text{peak}}$ at north pole) and radius $R_b$. We define the boundary of the bubble as the area where the perturbation amplitude falls to 1/10 of its peak value, the radius is then the distance between the peak and the boundary. Figure~\ref{fig:bubble_growth} plots the evolution of these two quantities. The bubble height grows in a manner qualitatively similar to the first order amplitudes, which is expected as the perturbation growth is dominated by first order. Concurrently, the bubble radius $R_b$ slowly decreases. This combined behavior with increasing height and decreasing radius indicates that the bubble sharpens as the evolution progresses.

\subsection{Influence of BP effects} \label{subsec:bp_results}

To isolate the effects of a static interface (i.e., suppressing the Bell–Plesset effects), we perform an additional numerical experiment in which the interface is held fixed and no convergent flow is present. Furthermore, the previously applied acceleration is replaced by a external force $g$ to give reasonable comparison. Such that, the radius is held constant, i.e., \(R = R_0\), while the driving force retains the same functional form \(g = \ddot R(t)\) from Eq.~\eqref{eq:implosion_presciption}, for the case \((l_0,m_0)=(4,\pm1)\).

\begin{figure}[htbp]
  \centering
  \includegraphics[width=1\linewidth]{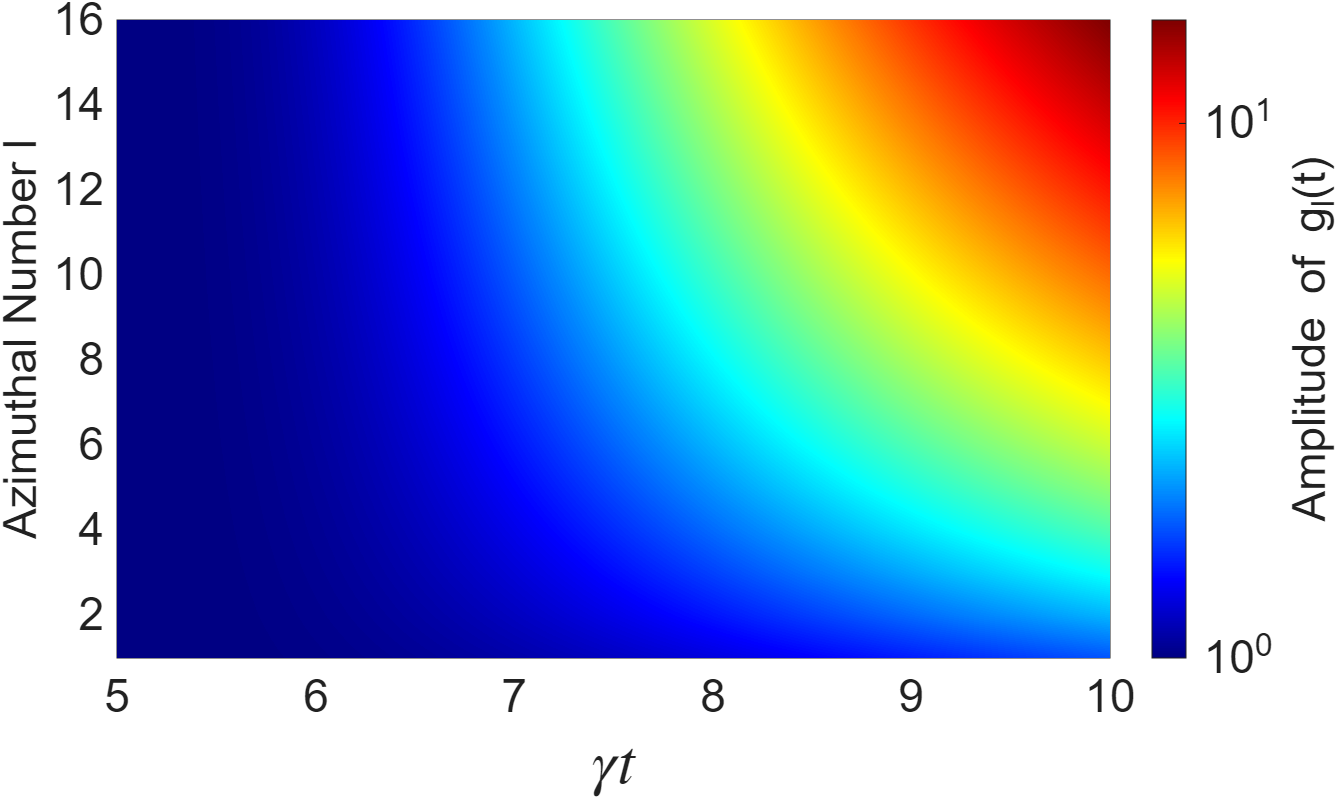}
  \caption{Heatmap of the first order growth factor \(g_l(t)\) for modes \(l = 1\) to \(20\), evolved from \(\gamma t = 5\) to 10 under the condition of a fixed interface radius (no BP effects).}
  \label{fig:first_order_gl_heatmap_noBP}
\end{figure}

Figure~\ref{fig:first_order_gl_heatmap_noBP} presents the heatmap of the first order amplitude growth factors under this static interface scenario. Compared to the corresponding plot in Figure~\ref{fig:first_order_gl_heatmap}, the pattern of mode behaviour is identical, but the amplitude is reduced by roughly two orders of magnitude. This clearly demonstrates that although the qualitative structure of the evolution remains, the presence of Bell–Plesset curvature/convergence effects drastically amplifies the growth rate of the instability.

\begin{figure}[htbp]
  \centering
  \includegraphics[width=1\linewidth]{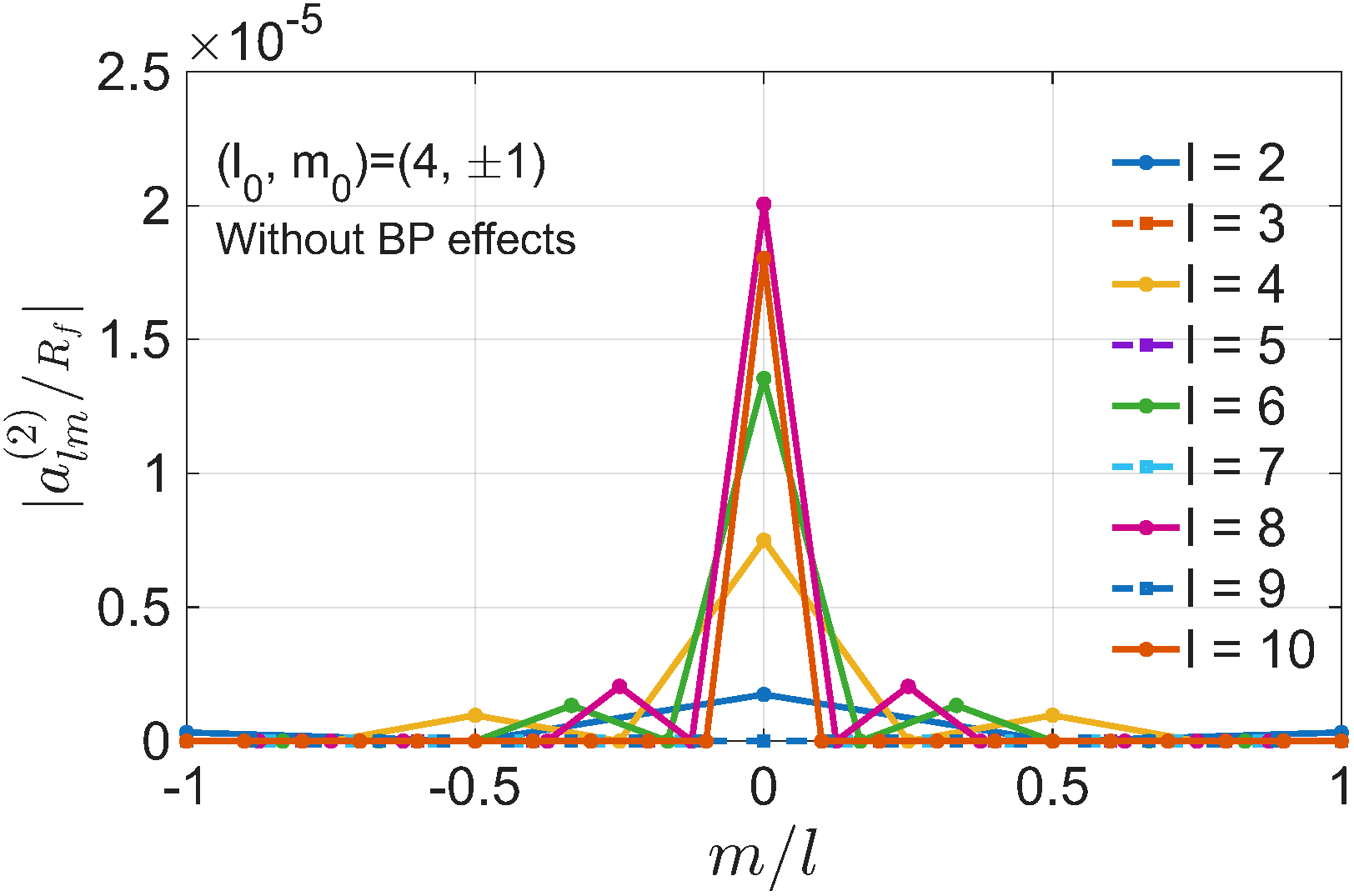}
  \caption{Spectrum of normalized second order amplitudes \(\bigl|a^{(2)}_{l,m}\bigr|/R_f\) for \(l = 2\) to 10 at \(\gamma t = 10\), for the initial condition $(l_0,m_0)=(4,\pm1)$, under the condition of a fixed interface radius (no BP effects).}
  \label{fig:spectrum_l4m1_noBP}
\end{figure}

Figure~\ref{fig:spectrum_l4m1_noBP} shows the second order amplitudes at the final time under the no‐BP scenario. While the shape of the spectrum is similar to that in Figure~\ref{fig:spectrum_all} (c), the magnitudes of the amplitudes are roughly three orders of magnitude smaller. This observation indicates that the BP effects exerts an even stronger influence on the growth of higher order mode couplings than it does on first order growth, underscoring the importance of incorporating BP effects in our analysis.

\section{Discussions and Conclusions} \label{sec:conclusions}

In this work, we developed a weakly nonlinear, multi-mode model to investigate the Rayleigh–Taylor instability (RTI) on a dynamic, three-dimensional spherical interface applicable to arbitrary initial conditons. Our model offers a comprehensive framework for analyzing RTI on a time-varying spherical surface, moving beyond many of the simplifying assumptions used in previous studies. Our analysis demonstrates that a complete description of RTI in such geometries must account for both the Bell–Plesset effects and nonlinear mode coupling, as these phenomena fundamentally alter the perturbation dynamics. The primary conclusion of our study is the identification of a powerful selection rule governing the nonlinear energy transfer between modes. We find that axisymmetric modes (those with azimuthal index $m=0$) preferentially gain energy from other modes, making this specific class of perturbations the most susceptible to rapid, destabilizing growth. This signature is fundamentally different from that observed in RTI without BP effects, where mode growth from a single mode input is largest near \(m \approx (l+1)/2\)~\cite{RTI_3d_spherical1}. The discrepancy arises because the driving mechanisms are distinct: in our dynamic case, the instability is driven by the time-dependent interface deceleration, whereas in the classic static case, it is driven by a constant gravitational field.

Moreover, our results show that for a modest compression ratio \(R_0/R_f = 5\), the instability amplitude can be amplified by an order of \(10^2\). This finding aligns with the scaling arguments presented by R. Epstein~\cite{BPe3}. Importantly, we further show that this amplification becomes progressively more pronounced at higher perturbation orders. The geometric convergence therefore has an even stronger impact on the growth of second and higher order mode couplings beyond the first order growth.

The theoretical framework presented here is robust and can be extended to other scenarios beyond the idealized implosion studied. For example, it is directly applicable to explosive events, or any arbitrary interface evolution. Future investigations could also incorporate more complex physics, such as time-dependent Atwood number and $\gamma_\rho$, and multi-mode initial conditions derived from experimental data, to explore their influence on the mode coupling dynamics. Expanding the spherical harmonic truncation to higher values of \(l_{\rm max}\) would allow inclusion of smaller-scale perturbations and yield more precise quantitative predictions.

Ultimately, the insights from this study have direct implications for several key scientific fields. In astrophysics, the preferential growth of low $l$, axisymmetric modes may help explain the large-scale mixing observed in core-collapse supernovae~\cite{SNe1,SNe2,SNe3,SNe4}. In ICF, our results highlight the critical importance of minimizing axisymmetric imperfections on capsule surfaces to control instability and achieve successful ignition~\cite{Hurricane1,Hurricane2,Hurricane3,Lawrence2023,Hurricane3}.

\section{Acknowledgments}
The authors gratefully acknowledge Professor Lifeng Wang and Dr. Jing Zhang for their insightful discussions and valuable feedback. We also extend our thanks to Zhiyuan College, Shanghai Jiao Tong University for providing a stimulating research environment that facilitated our investigation into this fascinating field.

\subsection{Conflict of interest}
The authors have no conflicts to disclose.

\appendix
\onecolumngrid

\section{Equations of Conservation of Mass and Momentum}
\label{app:equation_expansion}
We here demonstrate how Eqs.~\ref{eq:velocity_continuity} and \ref{eq:pressure_continuity} are expanded up to second order.  
For Eq.~\ref{eq:velocity_continuity}, we have
\begin{equation}
    \frac{\partial \eta}{\partial t}  
    + \frac{1}{r^2\sin^2\theta}\frac{\partial\eta}{\partial \varphi}\frac{\partial\phi^i}{\partial \varphi}
    + \frac{1}{r^2}\frac{\partial\eta}{\partial \theta}\frac{\partial\phi^i}{\partial \theta}
    -\frac{\partial\phi^i}{\partial r} 
    +\left(\dot{R} - \frac{\partial \psi}{\partial r} \right)  
    = 0, \quad \text{at } r=R+\eta.
\end{equation}

The surface deformation can be expressed as
\begin{equation}
    \eta=\sum_{l,m} a_{l,m}Y_{l,m}
    =\sum_{l,m} \left(\epsilon a_{l,m}^{(1)}+\epsilon^2a_{l,m}^{(2)}\right)Y_{l,m},
\end{equation}
and, as an example, the internal potential $\phi^{\mathrm{in}}$ takes the form
\begin{equation}
    \phi^{\mathrm{in}}
    =\sum_{l,m} 
    \left(\epsilon b_{l,m}^{\mathrm{in}(1)}+\epsilon^2b_{l,m}^{\mathrm{in}(2)}\right)
    \left(\frac{r}{R}\right)^l
    Y_{l,m}.
\end{equation}

Since both Eqs.~\ref{eq:velocity_continuity} and \ref{eq:pressure_continuity} are evaluated at $r=R+\eta$, it is crucial to note that after differentiating $\psi$ and $\phi^i$ with respect to $r$, $\theta$, and $\varphi$, all terms containing $r$ must be expanded using $r = R + \eta$ in order to obtain the complete $\epsilon$-expansion.  
To illustrate this procedure, consider $\partial_r \phi^{\mathrm{in}}$ in Eq.~\ref{eq:velocity_continuity}:
\begin{equation}
\begin{split}
    \frac{\partial \phi^{\mathrm{in}}}{\partial r}
    &=\sum_{l,m} \frac{l(\epsilon b_{l,m}^{\mathrm{in}(1)}+\epsilon^2b_{l,m}^{\mathrm{in}(2)})}{R}
      \left(\frac{r}{R}\right)^{l-1} Y_{l,m}
      =\sum_{l,m} \frac{l(\epsilon b_{l,m}^{\mathrm{in}(1)}+\epsilon^2b_{l,m}^{\mathrm{in}(2)})}{R}
      \left(1+(l-1)\frac{\eta}{R}\right) Y_{l,m}\\
    &=\sum_{l,m}\Bigg\{ \epsilon\!\left[ \frac{l}{R} b_{l,m}^{\mathrm{in}(1)}Y_{l,m} \right]
      +\epsilon^2\!\left[\frac{l}{R} b_{l,m}^{\mathrm{in}(2)}Y_{l,m}
      +\frac{l(l-1)}{R^2}\!\sum_{l_1,m_1}\!b_{l,m}^{\mathrm{in}(1)}a_{l_1,m_1}^{(1)}Y_{l,m}Y_{l_1,m_1} \right]\Bigg\}.
\end{split}
\end{equation}

All other terms are expanded analogously.  
Two additional useful relations are
\begin{equation}
    \frac{1}{r}
    \simeq \frac{1}{R}\!\left[1-\frac{\eta}{R}+\frac{\eta^2}{R^2}\right],
    \qquad
    \frac{1}{r^2}
    =\frac{1}{R^2(1+\eta/R)^2}
    \simeq \frac{1}{R^2}\!\left[1-\frac{2\eta}{R}+\frac{3\eta^2}{R^2}\right],
    \qquad
    \frac{\eta}{R} \ll 1.
\end{equation}

Following this procedure, both Eq.~\ref{eq:velocity_continuity} and Eq.~\ref{eq:pressure_continuity} can be fully expanded to second order.  
Upon projecting the resulting expressions onto a particular spherical harmonic mode, one obtains the first- and second-order relations summarized in Eqs.~\ref{eq:bin1_by_a1}, \ref{eq:bex1_by_a1}, \ref{eq:first_order_eq}, \ref{eq:second_order_eq1_in}, \ref{eq:second_order_eq1_ex}, and \ref{eq:second_order_eq2}.

\section{Angular Projections and Mode Coupling}
\label{app:second_order_equation}

To fully project Eq.~\ref{eq:velocity_continuity} and Eq.~\ref{eq:pressure_continuity} onto a specific spherical harmonic mode, we first multiply both equations by $\sin^2\theta$. The $\sin^2\theta$ term itself contains spherical harmonic components; therefore, after projection, several types of mode coupling arise. In this section, we define the corresponding coefficients $A_{l,m,l_1,m_1}$, $B_{l,m,l_1,m_1,l_2,m_2}$, $H_{l,m,l_1,m_1,l_2,m_2}$, and $\Theta_{l,m,l_1,m_1,l_2,m_2}$, which quantify the coupling among different angular modes.

Before giving the explicit forms of the coefficients, we simplify the mode coupling integrals as they typically take the form
\begin{equation}
\sum_{l_1, m_1, l_2, m_2}\int Y_{l,m}(\hat{\mathbf{r}}) Y_{l_1,m_1}(\hat{\mathbf{r}}) Y_{l_2,m_2}(\hat{\mathbf{r}})\, \mathrm{d}\Omega,
\label{eq:example_sum_Ylm_integral}
\end{equation}
which describe how initially independent spherical harmonic modes interact. Even a single-mode perturbation can excite additional modes at second order. These integrals can be conveniently expressed using the Wigner 3\(j\)-symbols~\cite{Wigner1993,Aquilanti2007}. The product of two spherical harmonics can be expanded as
\begin{equation}
\begin{split}
Y_{l,m} Y_{l_1,m_1}
= \sum_{l_2, m_2} & \sqrt{\frac{(2l+1)(2l_1+1)(2l_2+1)}{4\pi}}\, Y_{l_2, m_2}^{*}\times
\begin{pmatrix}
l & l_1 & l_2\\
0 & 0 & 0
\end{pmatrix}
\begin{pmatrix}
l & l_1 & l_2\\
m & m_1 & m_2
\end{pmatrix},
\label{eq:Ylm_products_by_wigner_3j}
\end{split}
\end{equation}
where the asterisk denotes complex conjugation and
$\begin{pmatrix} l & l_1 & l_2\\ m & m_1 & m_2 \end{pmatrix}$
is the Wigner 3\(j\)-symbol. Integrating over the solid angle gives the well-known identity
\begin{equation}
\begin{split}
\int Y_l^m(\hat{\mathbf{r}}) Y_{l_1}^{m_1}(\hat{\mathbf{r}}) Y_{l_2}^{m_2}(\hat{\mathbf{r}})\, \mathrm{d}\Omega
&= \sqrt{\frac{(2l+1)(2l_1+1)(2l_2+1)}{4\pi}} 
\begin{pmatrix}
l & l_1 & l_2\\
0 & 0 & 0
\end{pmatrix}
\begin{pmatrix}
l & l_1 & l_2\\
m & m_1 & m_2
\end{pmatrix}.
\label{eq:simplify_by_wigner_3j}
\end{split}
\end{equation}
The 3\(j\)-symbols vanish unless the following selection rules are satisfied:
\begin{equation}
|l_1-l_2| \leq l \leq l_1+l_2, \quad m = m_1 + m_2.
\label{eq:3j_nonvanishing_condition}
\end{equation}
This property greatly simplifies computation and clarifies the physical origin of multi-mode interactions.

Since
\begin{equation}
\sin^2\theta=\frac{2}{3}-\frac{4\sqrt{5\pi}}{15}Y_{2,0},
\end{equation}
the first term describes the projection of $Y_{2,0}$ onto $Y_{l,m}$. We define
\begin{equation}
\begin{aligned}
A_{l,m,l_1,m_1}
&=-\frac{4\sqrt{5\pi}}{15}\int Y_{l,m}^*Y_{l_1,m_1}Y_{2,0}\sin\theta\, \mathrm{d}\theta\, \mathrm{d}\varphi \\
&=-\frac{4\sqrt{5\pi}}{15}(-1)^m\int Y_{l,-m}Y_{l_1,m_1}Y_{2,0}\sin\theta\, \mathrm{d}\theta\, \mathrm{d}\varphi \\
&=-\frac{4\sqrt{5\pi}}{15}(-1)^m
\sqrt{\frac{(2l+1)(2l_1+1)\times5}{4\pi}}
\begin{pmatrix}
l & l_1 & 2\\
0 & 0 & 0
\end{pmatrix}
\begin{pmatrix}
l & l_1 & 2\\
-m & m_1 & 0
\end{pmatrix},
\end{aligned}
\label{eq:A_lm1}
\end{equation}
which is nonzero only when $|l_1-2|\leq l\leq l_1+2$ and $m_1=m$.

The $\frac{2}{3}$ factor implies that a projection of $\frac{2}{3}Y_{l_1,m_1}Y_{l_2,m_2}$ contributes to the target mode $Y_{l,m}$. Thus, we define
\begin{equation}
\begin{aligned}
B_{l,m,l_1,m_1,l_2,m_2}
&=\int Y_{l,m}^*(\hat{\mathbf{r}})Y_{l_1,m_1}(\hat{\mathbf{r}})Y_{l_2,m_2}(\hat{\mathbf{r}})\, \mathrm{d}\Omega \\
&=(-1)^m\sqrt{\frac{(2l+1)(2l_1+1)(2l_2+1)}{4\pi}}
\begin{pmatrix}
l & l_1 & l_2\\
0 & 0 & 0
\end{pmatrix}
\begin{pmatrix}
l & l_1 & l_2\\
-m & m_1 & m_2
\end{pmatrix}.
\end{aligned}
\end{equation}

Additionally, by interacting with $-\frac{4\sqrt{5\pi}}{15}Y_{2,0}$, another projection term arises,
\begin{equation}
\begin{aligned}
C_{l,m,l_1,m_1,l_2,m_2}
&=-\frac{4\sqrt{5\pi}}{15}(-1)^m\int Y_{l,-m}Y_{l_1,m_1}Y_{l_2,m_2}Y_{2,0}\sin\theta\, \mathrm{d}\theta\, \mathrm{d}\varphi \\
&=(-1)^m\sum_{l_3}A_{l_3,m_3,l_2,m_2}
\int Y_{l,-m}Y_{l_1,m_1}Y_{l_3,m_3}\sin\theta\, \mathrm{d}\theta\, \mathrm{d}\varphi \\
&=\sum_{l_3}A_{l_3,m_3,l_2,m_2}\, B_{l,m,l_1,m_1,l_3,m_3},
\end{aligned}
\end{equation}
where the intermediate mode $Y_{l_3,m_3}$ serves as a coupling bridge between $Y_{l_2,m_2}$ and $Y_{2,0}$.  
This recursive formulation allows efficient use of Wigner 3\(j\) algebra to simplify higher order products.

The total projection coefficient for $\sin^2\theta$ acting on $Y_{l_1,m_1}Y_{l_2,m_2}$ is then
\begin{equation}
H_{l,m,l_1,m_1,l_2,m_2}
=\frac{2}{3}B_{l,m,l_1,m_1,l_2,m_2}+C_{l,m,l_1,m_1,l_2,m_2}.
\end{equation}

The last type of projection to consider involves the term
$\partial_\theta Y_{l_1,m_1}\,\partial_\theta Y_{l_2,m_2}\,\sin^2\theta$,
which must be projected onto the target spherical harmonic mode $Y_{l,m}$. This projection defines the coefficient $\Theta_{l,m,l_1,m_1,l_2,m_2}$. To start with, we have
\begin{equation}
\frac{\partial Y_{l,m}}{\partial\varphi}=imY_{l,m}, \qquad
\frac{\partial Y_{l,m}}{\partial\theta}
=\frac{1}{2}e^{-i\varphi}G_1(l,m)\,Y_{l,m+1}
-\frac{1}{2}e^{i\varphi}G_2(l,m)\,Y_{l,m-1},
\end{equation}
where we have defined
\begin{equation}
G_1(l,m)=\sqrt{(l-m)(l+m+1)}, \qquad
G_2(l,m)=\sqrt{(l+m)(l-m+1)}.
\end{equation}

We then define
\begin{equation}
\begin{aligned}
\Theta_{l,m,l_1,m_1,l_2,m_2}
&=\int Y_{l,m}^*\,
\sin^2\theta\,
\bigg[\frac{1}{2}e^{-i\varphi}G_1(l_1,m_1)\,Y_{l_1,m_1+1}
-\frac{1}{2}e^{i\varphi}G_2(l_1,m_1)\,Y_{l_1,m_1-1}\bigg]\\
&\quad\times
\bigg[\frac{1}{2}e^{-i\varphi}G_1(l_2,m_2)\,Y_{l_2,m_2+1}
-\frac{1}{2}e^{i\varphi}G_2(l_2,m_2)\,Y_{l_2,m_2-1}\bigg]
\sin\theta\, \mathrm{d}\theta\, \mathrm{d}\varphi,
\end{aligned}
\end{equation}
which expands to
\begin{equation}
\begin{aligned}
\Theta_{l,m,l_1,m_1,l_2,m_2}
&= \frac{1}{4} \int Y_{l,m}^*\, \Bigg\{ \\
&\quad \sin^2\theta\, e^{-2i\varphi}\,
G_1(l_1,m_1)G_1(l_2,m_2)\,
Y_{l_1,m_1+1}Y_{l_2,m_2+1} \\
&\quad - \sin^2\theta\,
G_1(l_1,m_1)G_2(l_2,m_2)\,
Y_{l_1,m_1+1}Y_{l_2,m_2-1} \\
&\quad - \sin^2\theta\,
G_2(l_1,m_1)G_1(l_2,m_2)\,
Y_{l_1,m_1-1}Y_{l_2,m_2+1} \\
&\quad + \sin^2\theta\, e^{2i\varphi}\,
G_2(l_1,m_1)G_2(l_2,m_2)\,
Y_{l_1,m_1-1}Y_{l_2,m_2-1}
\Bigg\}\sin\theta\, \mathrm{d}\theta\, \mathrm{d}\varphi.
\end{aligned}
\end{equation}

Using the relations
\begin{equation}
\begin{aligned}
\sin^2\theta&=\frac{2}{3}-\frac{4\sqrt{5\pi}}{15}Y_{2,0} \\
Y_{2,2} &=\frac{1}{4}\sqrt{\frac{15}{2\pi}}\sin^2\theta e^{2i\varphi}
\Rightarrow e^{2i\varphi}\sin^2\theta=4\sqrt{\frac{2\pi}{15}}Y_{2,2}, \\
Y_{2,-2} &=\frac{1}{4}\sqrt{\frac{15}{2\pi}}\sin^2\theta e^{-2i\varphi}
\Rightarrow e^{-2i\varphi}\sin^2\theta=4\sqrt{\frac{2\pi}{15}}Y_{2,-2},
\end{aligned}
\end{equation}
we finally obtain
\begin{equation}
\begin{gathered}
\Theta_{l,m,l_1,m_1,l_2,m_2}
=-\frac{1}{4}G_1(l_1,m_1)G_2(l_2,m_2)H_{l,m,l_1,m_1+1,l_2,m_2-1}\\
-\frac{1}{4}G_1(l_2,m_2)G_2(l_1,m_1)H_{l,m,l_1,m_1-1,l_2,m_2+1}\\
\\+\sqrt{\frac{2\pi}{15}}G_2(l_1,m_1)G_2(l_2,m_2)D_{l,m,l_1,m_1-1,l_2,m_2-1}\\
+\sqrt{\frac{2\pi}{15}}G_1(l_1,m_1)G_1(l_2,m_2)E_{l,m,l_1,m_1+1,l_2,m_2+1},
\end{gathered}
\end{equation}
where
\begin{equation}
\begin{aligned}
D_{l,m,l_1,m_1,l_2,m_2}
&=\int Y_{l,m}^*Y_{2,2}Y_{l_1,m_1}Y_{l_2,m_2}\sin\theta\, \mathrm{d}\theta\, \mathrm{d}\varphi \\
&=\sum_{l_3}B_{l_3,m_3,l_2,m_2,2,2}\, B_{l,m,l_1,m_1,l_3,m_3},
\end{aligned}
\end{equation}
and
\begin{equation}
\begin{aligned}
E_{l,m,l_1,m_1,l_2,m_2}
&=\int Y_{l,m}^*Y_{2,-2}Y_{l_1,m_1}Y_{l_2,m_2}\sin\theta\, \mathrm{d}\theta\, \mathrm{d}\varphi \\
&=\sum_{l_3}B_{l_3,m_3,l_2,m_2,2,-2}\, B_{l,m,l_1,m_1,l_3,m_3}.
\end{aligned}
\end{equation}
These definitions enable the systematic decomposition of all angular products and derivative couplings in terms of Wigner 3\(j\)-symbols, providing both analytical clarity and computational efficiency.

\section{Matrix Form of Second Order Equations}
\label{app:second_order_matrices}

Following the index mapping~\ref{eq:k_lm_relation}, the matrix $U$ takes the form
\begin{equation}
U(k,k_1) = \frac{2}{3}\delta_{k}^{k_1} + A_{k,k_1},
\end{equation}
where $\delta_x^y$ denotes the Kronecker delta function, and $A_{k,k_1}$ represents $A_{l,m,l_1,m_1}$ for the sets of $(l_i,m_i)$ corresponding to $k_i$. 
We adopt this convention throughout the remaining derivations. 
Under this notation, the matrices $V_i$ and the vectors $Q_i$ and $P_i$ are expressed as

\begin{equation}
V_\text{in}(k,k_1) = -\frac{2}{3}\frac{l}{R}\delta_{k}^{k_1} - \frac{l_1}{R}A_{k,k_1},
\end{equation}

\begin{equation}
V_\text{ex}(k,k_1) = \frac{2}{3}\frac{l+1}{R}\delta_{k}^{k_1} + \frac{l_1+1}{R}A_{k,k_1},
\end{equation}

\begin{equation}
Q_\text{in}(k) = \frac{1}{R^2}\sum_{k_1,k_2}\Big\{ - m_1 m_2
        a^{(1)}_{k_1} b^{\text{in}(1)}_{k_2}B_{k,k_1,k_2}
        -\big[(3\dot{R} + R\gamma_\rho)a^{(1)}_{k_1} a^{(1)}_{k_2} + l_2(l_2-1)a^{(1)}_{k_1} b^{\text{in}(1)}_{k_2}\big] H_{k,k_1,k_2}
         +a^{(1)}_{k_1} b^{\text{in}(1)}_{k_2} \Theta_{k,k_1,k_2} \Big\},
\end{equation}

\begin{equation}
    Q_\text{ex}(k) = \frac{1}{R^2}\sum_{k_1,k_2}\Big\{ - m_1 m_2 a^{(1)}_{k_1} b^{\text{ex}(1)}_{k_2}B_{k,k_1,k_2} 
    -\big[(3\dot{R} + R\gamma_\rho)a^{(1)}_{k_1} a^{(1)}_{k_2} + (l_2+1)(l_2+2)a^{(1)}_{k_1} b^{\text{ex}(1)}_{k_2}\big] H_{k,k_1,k_2} 
        + a^{(1)}_{k_1} b^{\text{ex}(1)}_{k_2} \Theta_{k,k_1,k_2} \Big\},
\end{equation}

\begin{equation}
\begin{split}
    P_\text{in}(k) = \sum_{k_1,k_2} \Bigg\{&\frac{l_2}{R}a^{(1)}_{k_1} \dot{b}^{\text{in}(1)}_{k_2} - l_2^2 \frac{\gamma_R}{R}  a^{(1)}_{k_1} b^{\text{in}(1)}_{k_2}  + \frac{1}{R^2}\bigg( \dot{R}^2 - R\ddot{R} - \frac{\dot{\gamma}_\rho R^2}{2}\bigg) a^{(1)}_{k_1} a^{(1)}_{k_2} 
    \\&+ \frac{1}{R}\Big[ l_2(l_2 -1)\gamma_R - l_2(2\gamma_R + \gamma_\rho)\Big] a^{(1)}_{k_1} b^{\text{in}(1)}_{k_2} + \frac{(2\gamma_R + \gamma_\rho)^2}{2} a^{(1)}_{k_1} a^{(1)}_{k_2} 
         + \frac{l_1 l_2}{2R^2} b^{\text{in}(1)}_{k_1} b^{\text{in}(1)}_{k_2} \Bigg\}H_{k,k_1,k_2} 
         \\&+ \frac{1}{2R^2}\sum_{k_1,k_2} \bigg( {b^{\text{in}(1)}_{k_1} b^{\text{in}(1)}_{k_2} } \Theta_{k,k_1,k_2} -m_1m_2b^{\text{in}(1)}_{k_1}  b^{\text{in}(1)}_{k_2} B_{k,k_1,k_2} \bigg),
\end{split}
\end{equation}

\begin{equation}
    \begin{split}
        P_\text{ex}(k)=\sum_{k_1,k_2} \Bigg\{&-(l_2+1)^2 \frac{\gamma_R}{R}  a^{(1)}_{k_1} b^{\text{ex}(1)}_{k_2}
         - \frac{l_2+1}{R}a^{(1)}_{k_1} \dot{b}^{\text{ex}(1)}_{k_2} + \frac{1}{R^2}\bigg( \dot{R}^2 - R\ddot{R} - \frac{\dot{\gamma}_\rho R^2}{2}\bigg) a^{(1)}_{k_1} a^{(1)}_{k_2} \\&+ \frac{1}{R}\Big[ (l_2+1)(l_2 +2)\gamma_R + (l_2+1)(2\gamma_R + \gamma_\rho)\Big] a^{(1)}_{k_1} b^{\text{ex}(1)}_{k_2} 
         \\&+ \frac{(2\gamma_R + \gamma_\rho)^2}{2} a^{(1)}_{k_1} a^{(1)}_{k_2} + \frac{(l_1+1) (l_2+1)}{2R^2} b^{\text{ex}(1)}_{k_1} b^{\text{ex}(1)}_{k_2} \Bigg\}H_{k,k_1,k_2} \\
        & + \frac{1}{2R^2}\sum_{k_1,k_2} \bigg( b^{\text{ex}(1)}_{k_1} b^{\text{ex}(1)}_{k_2} \Theta_{k,k_1,k_2} -m_1m_2b^{\text{ex}(1)}_{k_1}  b^{\text{ex}(1)}_{k_2} B_{k,k_1,k_2} \bigg).
    \end{split}
\end{equation}

Here, $B_{k,k_1,k_2}$, $H_{k,k_1,k_2}$, and $\Theta_{k,k_1,k_2}$ respectively represent $B_{l,m,l_1,m_1,l_2,m_2}$, $H_{l,m,l_1,m_1,l_2,m_2}$, and $\Theta_{l,m,l_1,m_1,l_2,m_2}$ under the same mapping between $(l_i,m_i)$ and $k_i$.

\section{Convergence Test of Truncation Index $l_\mathrm{max}$}
\label{app:lmax_convergence}

\begin{figure}[htbp]
  \includegraphics[width=0.5\linewidth]{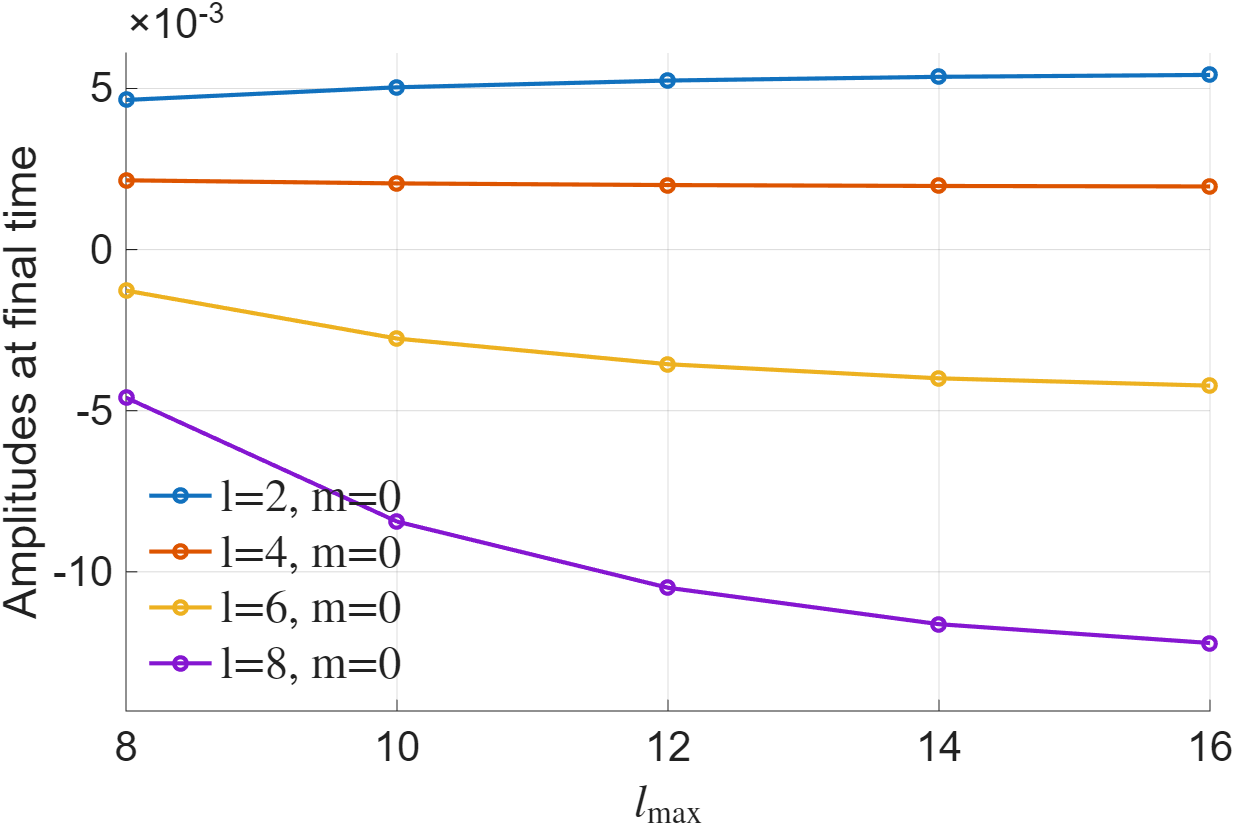}
  \caption{Convergence test of second-order amplitudes \(|a^{(2)}_{l,m}|\) for modes \(l = 2,4,6,8\) with \(m=0\) as a function of truncation index \(l_{\rm max}\) (ranging from 8 to 16). The vertical axis shows the amplitude \(a\) (normalized by \(R_f\)) and the horizontal axis shows \(l_{\rm max}\) from 8 to 16.}
  \label{fig:convergence_lmax}
\end{figure}

To determine an appropriate truncation for the spherical harmonic expansion, \(l_\text{max}\), we perform numerical tests using the \((l,m) = (4,\pm 1)\) mode as the initial condition and vary \(l_\text{max}\). Figure~\ref{fig:convergence_lmax} shows the second order amplitudes at the end of the evolution for several representative modes. The \(l=2\) and \(l=4\) modes converge rapidly at \(l_\text{max}\leq12\), while higher-\(l\) modes converge more slowly due to their proximity to the truncation limit. Overall, the low-$l$ modes that are of interests already converge at $l_\text{max}=16$.

\section{Dependence of Axisymmetric Tendency on $l_0$}
\label{app:axisymmetric_tendency}

\begin{figure}[htbp]
  \includegraphics[width=0.5\linewidth]{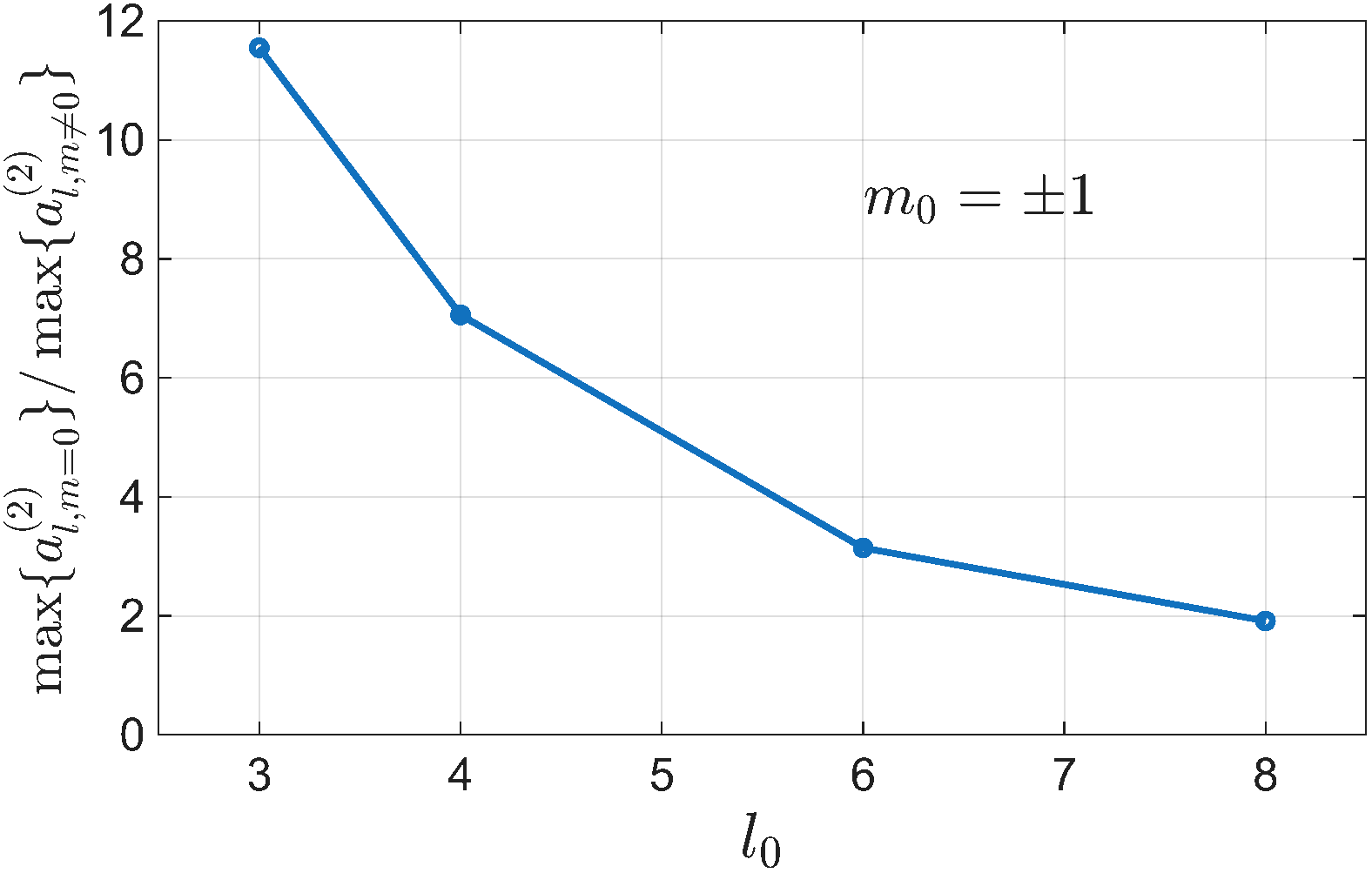}
  \caption{Ratio of the maximum axisymmetric mode to the maximum non-axisymmetric mode ($\max\{a^{(2)}_{l,m=0}\}/\max\{a^{(2)}_{l,m\neq 0}\}$) at $\gamma t=10$, as a function of the initial mode number $l_0$. The initial conditions consist of modes $(l_0, m_0) = (l_0, \pm 1)$ for $l_0 = 3, 4, 6, 8$. The decreasing trend indicates that the tendency toward axisymmetry becomes weaker at higher $l$ values.}
  \label{fig:axisymmetry_l_trend}
\end{figure}

To address whether the tendency toward axisymmetry becomes stronger or weaker at higher $l_0$ values, we performed a series of single-mode simulations with varying $l_0$, where $m_0$ is fixed to $\pm1$ and the initial amplitude is $\eta_0$. For each simulation, we quantify the relative growth of second-order axisymmetric modes by computing the ratio
\begin{equation}
    \mathcal{R}(l_0) = \frac{\max\{|a^{(2)}_{l,m=0}|\}}{\max\{|a^{(2)}_{l,m\neq 0}|\}}
    \label{eq:axisym_ratio}
\end{equation}
at the end of the evolution ($\gamma t=10$).

Figure~\ref{fig:axisymmetry_l_trend} shows $\mathcal{R}$ as a function of $l_0$. A clear decreasing trend is observed, indicating that higher-$l$ perturbations exhibit a weaker preference for axisymmetric growth. This suggests that the second-order coupling mechanism driving axisymmetric mode growth becomes less dominant at higher spherical harmonic degrees, possibly due to the increased number of non-axisymmetric channels available for energy transfer.

\section{Details of Bubble Behaviors}
\label{app:bubble_detail}

\begin{figure}[htbp]
  \centering
  \begin{subfigure}[b]{0.3\textwidth}
    \centering
    \includegraphics[width=\linewidth]{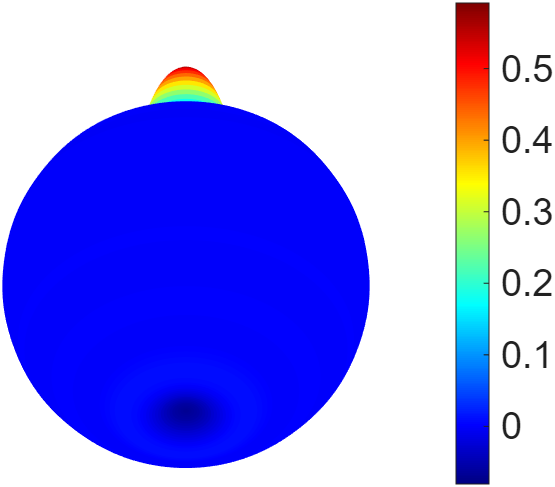}
    \caption{}
    \label{fig:bubble_north_show_bottom}
  \end{subfigure}
  \hfill
  \begin{subfigure}[b]{0.5\textwidth}
    \centering
    \includegraphics[width=\linewidth]{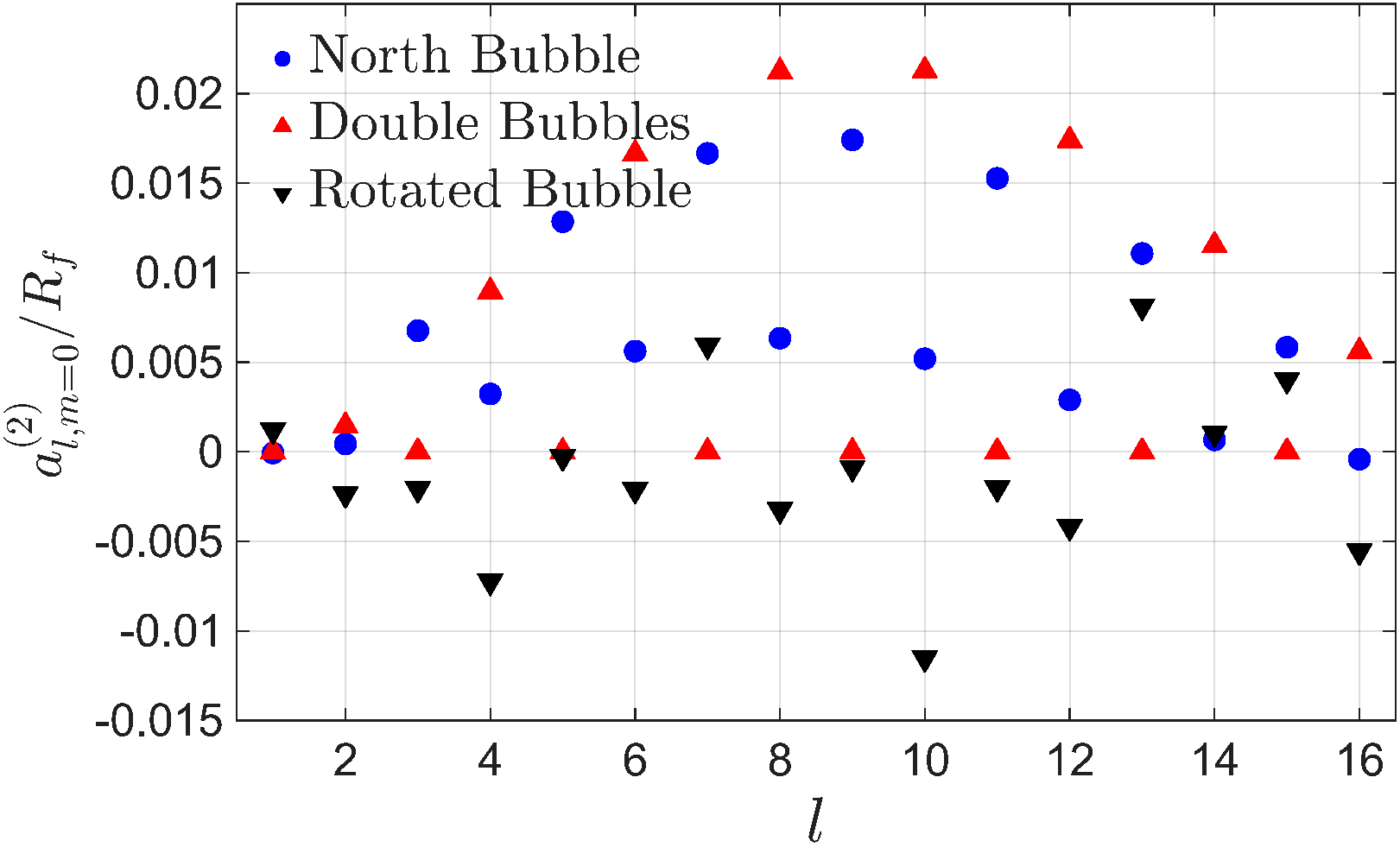}
    \caption{}
    \label{fig:bubble_detail}
  \end{subfigure}
  \caption{Additional analysis of bubble perturbations. (a) Bottom view of the evolved interface at $\gamma t = 10$ for the single bubble perturbation initialized at the north pole. The south pole region remains smooth and nearly spherical, with a small deformation of $\eta(\theta=\pi)/R_f=0.077$ due to the truncation error. (b) Second-order axisymmetric amplitudes ($a^{(2)}_{l,m=0}/R_f$) at $\gamma t = 10$ for the three bubble perturbation cases.}
  \label{fig:bubble_appendix}
\end{figure}

A potential concern with the spherical harmonic expansion is whether a localized perturbation at one pole might introduce spurious disturbances at distant regions of the interface. To verify that our method correctly captures the localized nature of the bubble perturbation, we examine the interface behavior at the south pole for the single-bubble simulation initialized at the north pole.

Figure~\ref{fig:bubble_north_show_bottom} shows the interface viewed from below at $\gamma t = 10$. The south pole region remains smooth and nearly spherical, despite showing a small deformation due to the truncation error. This confirms that the spherical harmonic expansion faithfully preserves the localized character of the perturbation, without introducing artificial disturbances in regions far from the initial perturbation site.

Figure~\ref{fig:bubble_detail} shows the real values of the second-order amplitudes for $m=0$ modes. For the bubbles located at the north and south poles, the amplitudes are predominantly positive, whereas for the rotated bubble case, they alternate between positive and negative values, revealing the rich structure inherent in the second-order evolution.

\twocolumngrid

\bibliography{sections/references.bib}

\end{document}